\documentclass[cmp]{svjour}   
\usepackage{amssymb}
\renewcommand\a{\"a}
\newcommand\C{\mathbb{C}}
\newcommand\E{\mathbb{E}}
\newcommand\F{\mathbb{F}}

\renewcommand\P{\mathbb{P}}
\newcommand\Q{\mathbb{Q}}
\newcommand\R{\mathbb{R}}

\renewcommand\S{\mathbb{S}}
\newcommand\Z{\mathbb{Z}}
\renewcommand\j{\jmath} 
\newcommand\diag{\mathop{\mathrm{diag}}}
\newcommand\mod{\mathop{\mathrm{mod}}}
\renewcommand\span{\mathop{\mathrm{span}}\nolimits}

\newcommand\inv[2][1]{ {\textstyle {#1\over #2}} }
\def\id{\mathchoice{\setlength{\unitlength}{1ex}\identitaet}{
\setlength{\unitlength}{1ex}\identitaet}{
\setlength{\unitlength}{0.5ex}
\begin{picture}(1.6,1.5)
\put(0.7,-0.1){\scriptsize 1}\thinlines
\put(1,0.1){\line(0,1){1.4}}
\put(0.2,0){\line(1,0){1.2}}
\put(0.6,1.5){\line(1,0){0.4}}
\end{picture}}{\setlength{\unitlength}{0.5ex}
\begin{picture}(1.6,1.5)
\put(0.7,-0.1){\scriptsize 1}\thinlines
\put(1,0.1){\line(0,1){1.4}}
\put(0.2,0){\line(1,0){1.2}}
\put(0.6,1.5){\line(1,0){0.4}}
\end{picture}}
}
\newcommand\identitaet{\begin{picture}(1.6,1.5)
\put(0,0){1}
\put(1,0.1){\line(0,1){1.4}}
\thinlines
\put(0.2,0){\line(1,0){1.2}}
\put(0.6,1.5){\line(1,0){0.4}}
\end{picture}}
\newcommand\ds{\displaystyle}

\begin{document}
\title{Mirror Symmetry on Kummer Type $K3$ Surfaces} 
\author{Werner Nahm \and Katrin Wendland 
}
\titlerunning{Mirror Symmetry on Kummer Type $K3$  Surfaces}
\author{
Werner Nahm\inst{1} \and
Katrin Wendland\inst{2}
}
\institute{Physikalisches Institut, Universit\a t Bonn,
Nu\ss allee 12, D-53115 Bonn, Germany.\\
\email{werner@th.physik.uni-bonn.de}
\and
Dept. of Physics and Astronomy, UNC at Chapel Hill,
141 Phillips Hall, CB \#3255, Chapel Hill, N.C. 27599.
\email{wendland@physics.unc.edu}}
\date{}
\authorrunning{Werner Nahm and Katrin Wendland}
%
%
%
%
\maketitle
\begin{abstract}
We investigate both geometric and conformal field theoretic aspects of
mirror symmetry
on $N=(4,4)$ superconformal field theories with central charge $c=6$.
Our approach enables us to
determine the action of mirror symmetry
on (non-stable) singular fibers in elliptic fibrations
of $\Z_N$ orbifold limits of $K3$.
The resulting map gives an automorphism of   order $4,8$, or $12$,
respectively, on the smooth universal
covering space of the moduli space. We  explicitly derive the
geometric counterparts of the twist fields in our orbifold conformal
field
theories. The classical McKay correspondence allows for a natural
interpretation of
our results.
\end{abstract}
\section{Introduction}
We investigate the version of mirror symmetry \cite{di87,lvw89,grpl90}
which was found  by Vafa and Witten for 
orbifolds of   toroidal theories   \cite{vawi95},  and which
was generalized to the celebrated Strominger/Yau/Zaslow conjecture
\cite{syz96}.
Since the conceptual issues of mirror symmetry for
$N=(4,4)$ superconformal field theories 
on first sight are different
and more controversial than for strict $N=(2,2)$ theories,
we first discuss the latter, as a preparation.

An $N=(2,2)$ superconformal field theory
is a fermionic conformal field theory
(CFT)   together with a marking, i.e.\ a map
from the standard super--Virasoro algebra  into the 
operator product expansion (OPE) of this theory. 
Due to the marking the theory has well defined
left and right handed $U(1)$ charges $Q_l , Q_r$. Markings
which differ by $Q_l, Q_r$ gauge transformations are identified. 
Results on $N=(2,2)$ deformation theory \cite{di87} show that 
for given central charge $c$ a moduli
space of $N=(2,2)$ superconformal field theories can be defined.
Its irreducible components at generic points are   Riemannian
manifolds under the Zamolodchikov metric \cite{za86} and have at most
orbifold singularities. Note  that the completion of the moduli space 
may contain
points of extremal transitions which do not have CFT descriptions and will
not be of relevance for our discussion.
We shall
restrict considerations to a connected 
part  of the moduli space. 
By the above it has a unique
smooth, simply connected
covering space $\widetilde{\mathcal M}$ \cite{th97}.
We also assume that all 
theories in our moduli space 
include the spectral flow operators in their Hilbert
spaces. 
 
Let us now consider
strict $N=(2,2)$ theories, i.e.\ those for which
the marking has no continuous deformations. 
Then with respect to the left and right $U(1)$ action the
tangent bundle $T\widetilde{\mathcal{M}}$ canonically  splits into
two subbundles. 
The cover of the moduli space
has a corresponding canonical product realization
$\widetilde{\mathcal M}=\widetilde{\mathcal M}_1\times
\widetilde{\mathcal M}_2$ \cite{di87,digr88}.
Let $\Gamma^0\subset \Gamma$ be the subgroup of elements
which admit a factorization
$\gamma=\gamma_1\gamma_2$ such that  $\gamma_i\in\Gamma$ only 
acts on $\widetilde{\mathcal M_i}$ (other automorphisms may exist that are 
related to the effect of monodromy \cite{ghl96,kmp96}).
The factorization of $\widetilde{\mathcal M}$ induces a
factorization
$\mathcal M:=\widetilde{\mathcal M}/\Gamma^0
= {\mathcal M}_1\times {\mathcal M}_2$.
The corresponding two subbundles of $T{\mathcal M}$
are distinguished by the marking.
The  standard mirror automorphism of the
super--Virasoro OPE which inverts the sign of one of the $U(1)$ generators  
interchanges these subbundles.

One expects that near some boundary component of the moduli space
any of our theories has a geometric interpretation as  supersymmetric
sigma model on a space $X$ with Ricci flat K\a hler metric of large
radius. Though there may be several boundary components of this kind which
yield manifolds that cannot be deformed into each other as algebraic
manifolds, but are birationally equivalent manifolds 
\cite{agm94b,ko95b,delo99,ba99}, 
we choose a unique $X$ for ease of exposition,
and we write $\mathcal M=\mathcal M(X),
\widetilde{\mathcal M}=\widetilde{\mathcal M}(X)$. 
One possible
deformation of $X$ is given by the scale transformation of the metric.
It turns out to
belong to the tangent space of one of the factors,  say
$\mathcal M_1$. Then $\mathcal M_2$ becomes the space of complex
structures on $X$, and close to the boundary $\mathcal M_1$ corresponds
to variations of the (instanton corrected) complexified K\a hler structure.
Under mirror symmetry the roles of the two factors
are interchanged, such that $\mathcal M_1$ becomes the moduli space
of complex structures on some other space $X^\prime$.
This induces a duality between geometrically different
Calabi Yau manifolds, an observation that has had a striking impact
on both mathematics and physics (see \cite{cogp91,mo93}, and
\cite{coka99} for a more complete list of references).

Let $\Sigma(X)$ be the space of sigma model Lagrangians on $X$
(possibly with a marking of the homology of $X$), and
$\Sigma(X)_b\subset \Sigma(X)$ a connected 
and simply connected boundary region
whose points define a conformal field theory by some quantization
scheme. Let 
$\sigma_b(X): \Sigma(X)_b\rightarrow {\mathcal M}(X)$
be the corresponding continuous map. 
Locally, $\sigma_b(X)$ is a homeomorphism.
By deformation, a mirror symmetry
$\Sigma(X)_b\cong\Sigma(X^\prime)_b$
induces an isomorphism
${\mathcal M}(X)\cong{\mathcal M}(X^\prime)$. Since isomorphic
CFTs yield the same point in ${\mathcal M}(X)$, such an 
isomorphism cannot depend on the choice of the boundary region.
When $X=X^\prime$ the induced automorphism of ${\mathcal M}(X)$
corresponds to an automorphism of the CFT which changes
the sign of one of the two $U(1)$ currents and exchanges the
corresponding supercharges.
This automorphism changes the marking and therefore acts non-trivially
on ${\mathcal M}(X)$.

When a base point has been chosen, the local isomorphism 
$\sigma_b(X)$ lifts to an inclusion
$\widetilde\sigma_b(X): \Sigma(X)_b\hookrightarrow 
\widetilde{\mathcal M}(X)$ and analogously for $X^\prime$.
When $X=X^\prime$ and $\Sigma(X)_b\cap \Sigma(X^\prime)_b$
is non-empty, we choose a base point in the intersection.
Then $\sigma^\prime_b(X^\prime)\circ \sigma_b(X)^{-1}$ lifts to a
mirror isomorphism
$\gamma_{ms}^b: \widetilde{\mathcal M}(X)\rightarrow
\widetilde{\mathcal M}(X^\prime)$.
Sometimes there are canonical choices for the base point,
otherwise one obtains equivalent choices which correspond
to multiplication of $\gamma_{ms}^b$ by an element of some
subgroup of $\Gamma$.

Given a Calabi-Yau manifold
$X$ one can construct the corresponding
family of sigma models,
the moduli space $\mathcal{M}(X)$, and the mirror Calabi-Yau manifold
$X^\prime$.
Since $X,X^\prime$ determine   (families of)
classical geometric objects, it should be possible
to transform one into the other by purely classical methods.
Such constructions for mirror pairs have been proposed in the context
of toric geometry \cite{ba94} as well as T--duality
\cite{vawi95,syz96,koso00}.

The basic case is given by   theories with central charge $c=3$,
which correspond
to sigma models on a two-dimensional torus. Here one has
$\widetilde{\mathcal M}=\mathcal{H}\times \mathcal{H}$, where
we shall use   coordinates $\rho,\tau$ for the two copies of the
upper complex half-plane
$\mathcal{H}\cong\widetilde{\mathcal M}_1\cong\widetilde{\mathcal M}_2$.
In our conventions on automorphisms, the fundamental group is
given by the standard $SL(2,\Z)\times SL(2,\Z)$ action on
$\widetilde{\mathcal M}$.
Orientation change, given by $\rho\mapsto -\bar\rho$,
$\tau\mapsto -\bar\tau$, and space parity change, given by
$\rho\mapsto -\bar\rho$, $\tau\mapsto \tau$, are not considered
as automorphisms, since they change the marking.
For purely imaginary $\rho,\tau$
the theory is the (fermionic) product of circle models with squared radii
$r_1^2=\rho/\tau$, $r_2^2=-\rho\tau$. Hence our theory has a nonlinear sigma
model description given by two Abelian $U(1)$ currents $j_1,j_2$, and their
$N=2$
superpartners $\psi_1,\psi_2$, together with analogous right handed fields
$\overline\j_1,\overline\j_2, \overline\psi_1,\overline\psi_2$, 
all compactified on a real torus. The
$\psi_i,\overline\psi_i$ are Majorana fermions. 
The eigenvalues of the four currents $j_i, \overline\j_i$
lie in a four-dimensional vector space
with natural $O(2)_{left}\times O(2)_{right}$ and $O(2,2)$ actions. Due
to the compactification they form a lattice of rank $4$.
The $U(1)$ current of the left-handed
$N=2$ superconformal algebra is 
$J=$ \mbox{$i\!:\!\!\psi_2\psi_1\!\!:$}. Hence mirror symmetry can be induced
by the OPE preserving map which leaves right handed fields unchanged and 
transforms left handed ones by
$(\psi_1,\psi_2)\mapsto(-\psi_1,\psi_2)$,
$(j_1,j_2)\mapsto(-j_1,j_2)$. For the fermionic sigma model on the first
circle this is the T--duality map, i.e.\ $r_1\mapsto (r_1)^{-1}$. 
The existence of this OPE preserving map implies that
$\mathcal M_1$ is isomorphic to $\mathcal M_2$, as stated above. 
Summarizing, close to the ``boundary
point'' $\tau=i\infty,\rho=i\infty$ of $\widetilde{\mathcal M}$
mirror symmetry is given by the
exchange of $\rho$ and $\tau$. This is fiberwise T-duality in an
$\S^1$ fibration (with section) on the underlying torus
and motivates the construction \cite{vawi95,syz96}.
The cusp $\tau=i\infty,\rho=i\infty$ corresponds to a limit
where the base volume $r_2$ becomes infinite, whereas the 
relative size of the fiber is arbitrarily small, e.g.\ for constant
$r_1$. Since base space and fiber are flat, semiclassical considerations
are applicable for arbitrary $r_1$, such that mirror symmetry yields a
relation between two classical spaces.  
Conjecturally the idea carries over to suitable torus fibrations over more
complicated base spaces of infinite volume. 
Mathematically, the expected map is given by a Fourier--Mukai type functor
\cite{ko95,mo96}.

The construction of mirror symmetry by fiberwise T-duality
also makes sense when $X$ is a 
hyperk\a hler manifold and the corresponding sigma model has
a superconformal symmetry which is extended beyond $N=(2,2)$,
though the relationship with CFT is somewhat different.
The moduli space $\mathcal{M}$ no longer splits canonically,
since there is no canonical $N=(2,2)$ subalgebra of the extended super
Virasoro algebra. Moreover, there
are no quantum corrections to the K\a hler structure, such that each
point of $\widetilde{\mathcal{M}}$ corresponds to a classical sigma
model, with well defined Ricci flat metric and B-field.
In this situation classical geometries corresponding
to different boundary components should be diffeomorphic,
since for compact hyperk\a hler manifolds this is implied by birational
equivalence \cite{hu99}.
According to the previous arguments, mirror symmetry must yield
an element $\gamma_{ms}$ of $\Gamma$, which depends on the geometric
interpretation.

The picture developed so far is conjectural, but in some cases it             
can be verified, since the moduli space is entirely known. This is true
for toroidal theories and for those  theories with $c=6$ whose Hilbert 
spaces include the spectral flow operators
\cite{na86,asmo94,nawe00}. Every such theory
admits geometric interpretations in terms of nonlinear sigma models 
either on tori or on $K3$, depending on the CFT. Thus any 
mirror symmetry relates two different geometric interpretations 
within the same moduli space $\mathcal{M}$.

In this note,   we explore a version of mirror symmetry  
on Kummer type $K3$ surfaces  that was proven
in a much more general context in \cite{vawi95} and actually led to the
Strominger/Yau/Zaslow conjecture, see
\cite{syz96,grwi97,mo96,ven00}.
Namely, for a $T^2$ fibered $K3$ surface $p:X\longrightarrow\P^1$
with elliptic fibers and a section,
mirror symmetry is induced by T--duality on each regular fiber of $p$.
Among various maps known as mirror symmetry,
this is the only one with general applicability.  
We show that it generalizes to the singular fibers and determine the
induced map. 
It turns out to be of finite order $4,8$, or $12$ in the  different cases we 
discuss.
Note that by construction, our mirror map depends on the respective 
geometric interpretations on orbifold limits of $K3$. In other words,
we are forced to work on the cover $\widetilde{\mathcal{M}}$ of the moduli
space. It would be interesting to understand the precise effect of
quotienting out by the automorphism group of $\widetilde{\mathcal{M}}$.
Our approach enables us to read off 
the exact identification of twist fields in the relevant orbifold
conformal field theory with geometric data on the corresponding Kummer
type $K3$ surface. 
The role of ``geometric'' versus ``quantum'' symmetries is thereby
clarified.
The correct identification is also of major importance for
the discussion of orbifold cohomology and resolves the objection of
\cite{fago01} to Ruan's conjecture \cite{ru00}
on the orbifold cohomology
for hyperk\a hler surfaces\footnote{\label{ruan}The 
fact that our transformation
resolves this objection was explained to us by Yongbin Ruan 
\cite{ru01} and goes
back to an earlier observation by Edward Witten.}. For those Kummer type
$K3$ surfaces discussed in this note, our results in fact prove part of 
Ruan's conjecture.

To make the paper more accessible to mathematicians, we do not
use the language of branes for the geometric data, but a translation
is not hard.

This work is organized as follows:
In Sect.~\ref{tori} we discuss our mirror map on four-tori. In 
Sect.~\ref{kummer} we show how this map induces a mirror map on Kummer
$K3$ surfaces $X$. In particular, we determine the induced map on the
(non-stable) singular fibers of our elliptic
fibration $p:X\longrightarrow\P^1$. In 
Sect.~\ref{kummertype} we give its generalization to other Kummer type
surfaces, i.e.\ other non-stable singular fibers. 
Sect.~\ref{twist}   deals with the CFT side of the picture: As 
explained above, the mirror map is an automorphism on a given superconformal
field theory. Its action on the bosonic part of the
Hilbert space of $\Z_N$ orbifold conformal
field theories on $K3$, $N\in\{2,3,4,6\}$, is determined independently
from the results of Sects.~\ref{kummer} and \ref{kummertype}. In 
Sect.~\ref{mckay} we use the previous results to read off an explicit
formula that maps twist fields to cohomology classes on $K3$. This is 
interpreted in terms of the
classical McKay correspondence.
We close with a summary and discussion in Sect.~\ref{conc}.
\subsection*{Acknowledgments:}
It is a pleasure to thank Paul Aspinwall, David R. Morrison,
M. Ronen Plesser, Miles Reid, Daniel Roggenkamp, Yongbin Ruan, 
Eric Sharpe, and Bernardo Uribe for helpful discussions.
W.N. thanks the LPTENS in Paris for an invitation, which led to the start of
part of this work. 
K.W. was supported by U.S. DOE grant DE-FG05-85 ER 40219, TASK A.
\section{The Mirror Map for Tori}\label{tori}
As pointed out in the Introduction, for an
$N=(2,2)$ superconformal field theory on a
two--dimensional orthogonal
real torus with radii $r,r^\prime$ and vanishing B--field, at large 
$r^\prime$, mirror symmetry  
is just the T--duality map
$r\mapsto r^{-1}$ for one radius, whereas $r^\prime$ remains unchanged. 
This map is naturally continued to arbitrary values
of $r,r^\prime$ \cite{vawi95}.  

Now consider a toroidal theory on  the Cartesian product $T$ of two 
two-di\-men\-sio\-nal orthogonal
tori with radii $r_1,r_3$, and $r_2,r_4$, respectively. Since the $U(1)$ 
currents of the $N=2$ superconformal algebras 
in the lower dimensional theories add up to give the $U(1)$ current of the 
full theory, mirror symmetry is induced by $r_1\mapsto (r_1)^{-1},
r_2\mapsto (r_2)^{-1}$ \cite{vawi95}. 
After a suitable choice of complex structure 
this is fiberwise T--duality on a special Lagrangian   
fibration (with section)
of our four--torus.
Hence we are discussing the version of mirror symmetry that was
generalized in \cite{syz96}, see also \cite{mo96,ven00}.
Alternatively, the fibration can be understood in terms of a
Gromov-Hausdorff collapse \cite{koso00}.

We wish to determine the corresponding map on the cover of the moduli space. 
Recall \cite{egta88} that in the present case of $N=(2,2)$ superconformal
field theories with central charge $c=6$ we actually have extended, i.e.\
$N=(4,4)$ supersymmetry. By \cite{na86,se88,ce90,asmo94,nawe00},
a theory in the corresponding moduli space is specified by the relative
position of an even self-dual lattice $L$ and a positive 
definite\footnote{On cohomology, we generally use the scalar product 
(\ref{intersection}) that is induced by
the intersection form on the respective surface.} 
four-plane $x$ in
$\R^{4,4+\delta}\cong H^{even}(Y,\R)$ with $\delta=0$ or $16$, depending
on whether the theory is associated to a torus or a $K3$ surface $Y$. 
In terms of parameters $(g,B)$ of nonlinear sigma models on $Y$, $g$
an Einstein metric on $Y$ and $B$ a B-field, and
for vanishing B--field, $x$ is 
the positive eigenspace of the
Hodge star operator $\ast$ in $H^{even}(Y,\R)$, and
$L=H^{even}(Y,\Z)$.
The symmetry group of $x$ is $SO(4)\times O(4+\delta)$, such that
\begin{equation}\label{cover}
\widetilde{\mathcal M} 
= O^+(4,4+\delta;\R)/\left(SO(4)\times O(4+\delta)\right),
\end{equation}
which is indeed simply connected \cite{wo65}.
We remark that for $\delta=16$, the space $\widetilde{\mathcal M}$ 
as in (\ref{cover}) is a partial completion of the smooth universal
covering space of the actual moduli space of $N=(4,4)$ SCFTs on $K3$. 
Namely, $\widetilde{\mathcal M}$ contains points which do not correspond
to well-defined SCFTs \cite{wi95}. They form subvarieties of 
$\widetilde{\mathcal M}$  with at least complex codimension one
\cite{agm94a}. These ill-behaved theories, however, will not be of relevance
for the discussion below.

For the torus we have $\delta=0$, and
the denominator in (\ref{cover}) contains
$SO(4)_{left}\times SO(4)_{right}$, the elements of which act as rotations
on  the left and right handed charges $Q_l,Q_r$, 
analogously to the case of the 
two-dimensional torus  discussed in the Introduction.
The fundamental group of the moduli space is given by
$$
\Gamma=Aut(L(Y_0))\cong O^+(4,4+\delta;\Z),
$$
where $L(Y_0)$ describes the base point.

To determine the element of $\Gamma$ that acts as mirror symmetry, it is 
crucial to gain a detailed understanding of the map that associates a
point in moduli space to given nonlinear sigma model data. 
In $d$ dimensions, it is customary to specify a toroidal theory by a lattice
with generators $\lambda_1,\dots,\lambda_d\in\R^d$, i.e.\
$T=\R^d/\span_\Z\{\lambda_1,\dots,\lambda_d\}$,
and a B-field $B\in Skew(d)$. Here $\R^d$ carries the standard metric.
We choose a reference torus $T_0$ given by the lattice
$\Z^d$ with standard orthonormal generators
$e_1,\dots,e_d$, and   $\Lambda\in Gl^+(d)$ such that 
$\lambda_i=\Lambda e_i$, $i\in\{1,\ldots,d\}$.

The group $Gl^+(d)$ has a natural representation on the dual space
$\check\R^{d}$ which is identified with $\R^d$ by the standard metric. 
The corresponding image of $\Lambda$ is
$$
M:=(\Lambda^T)^{-1}.
$$
The vectors $\mu_i:=Me_i,i\in\{1,\dots,d\}$, form a dual basis with respect
to $\lambda_1,\dots$ $\dots,\lambda_d$.
Similarly, we have a natural representation on $\Lambda^n(\check\R^{d})$.
The image of $\Lambda$ under this representation will be denoted
$\Lambda^n(M)$. Note that  $\Lambda^d(M)$ acts 
by multiplication with $V^{-1}$, where $V=\det(\Lambda)$
is the volume of the torus.  

In the standard description, the charge lattice of the theory is given
by pairs ${1\over\sqrt2}(Q_l+Q_r,Q_l-Q_r)\in \R^{d,d}$,
where it is natural to take 
$\R^{d,d}=\check\R^{d}\oplus \R^d\cong H^1(T_0,\R)\oplus H_1(T_0,\R)$ with  
bilinear form
\begin{equation}\label{bil}
(\alpha,\beta)\cdot(\alpha^\prime,\beta^\prime)
= \alpha\beta^\prime+\alpha^\prime\beta.
\end{equation}
The charge lattice is even and integral and is obtained as  image of the
standard lattice
$\Z^{d,d}\cong H^1(T_0,\Z)\oplus H_1(T_0,\Z)$
under 
\begin{equation}
v(\Lambda,B) =
\left( \begin{array}{cc}M & 0\\
                        0 & \Lambda\end{array}\right)
\left( \begin{array}{cc} \id & -B\\
                           0 & \id \end{array}\right)
\in O^+(d,d).
\end{equation} 
Here $B$ appears as a skew symmetric linear transformation from
$\R^d$ to $\check\R^{d}$. The corresponding element of
$\Lambda^2(\check\R^{d})$ will be denoted $b$, such that $b$ is
a vector with components $b_{ij}=B_{ij}$ with respect to the 
basis $e_i\wedge e_j$. 
We will also use its
dual $\check b$ with components $\check b_{ij}=
\sum_{k,l}\epsilon_{ijkl}B_{kl}/2$. 

As mentioned below (\ref{cover}), the rotations of the left and right
charge  lattices form a
subgroup $SO(d)_{left}\times SO(d)_{right}\subset O^+(d,d)$.
Rotations which
act on $Q_l$ and $Q_r$ in the same way are generated by
$$
\left( \begin{array}{cc} \Omega & 0\\
                          0  & \Omega \end{array}\right),
\quad\quad\Omega\in so(d).
$$ 
Rotations for which the respective actions on
$Q_l,Q_r$ are inverse to each other are generated by
$$
\left( \begin{array}{cc} 0  & \Omega\\
                          \Omega & 0 \end{array}\right),
\quad\quad\Omega\in so(d).
$$
To describe torus orbifolds we have to work with the lattice
$H^{even}(T_0,\Z)$ instead of $H^1(T_0,\Z)\oplus H_1(T_0,\Z)$. The
vector space $H^{even}(T_0,\R)$ carries 
a bilinear form $\langle\cdot,\cdot\rangle$ that we obtain
from the intersection form upon Poincar\'e duality, i.e.\
\begin{equation}\label{intersection}
\forall\;a,b\in H^{even}(T_0,\R): \quad
\langle a,b\rangle =\int_{T_0} a\wedge b.
\end{equation}
Accordingly, we have to
use a half spinor representation $s$ of $O^+(d,d)$. 
We now specialize
to the case $d=4$, where  
$H^1(T_0,\R)\oplus H_1(T_0,\R)\cong H^{even}(T_0,\R)$.
Hence the representation $s$ can be obtained from
$v$ by triality \cite{di99,nawe00,kop00,obpi01}. 
In other words,
\begin{equation}\label{spinor}
s(\Lambda,B) =
V^{1/2}\left( \begin{array}{ccc} V^{-1}   & 0  & 0\\
                                 0 & \Lambda^2(M) & 0\\
                                 0 &    0   & 1 \end{array}\right)
\left( \begin{array}{ccc}   1 & \check b & \;- {\|B\|^2\over2}\\
                           0 & \id & -b\\
                           0 & 0 & 1 \end{array}\right).
\end{equation}
Here $\|B\|^2=\langle B,B\rangle$ as given in (\ref{intersection}),
and by $\check b$ we denote the dual of $b$ as introduced above.
The matrix $s(\Lambda,B)$  acts on
$L(T_0):=H^{even}(T_0,\Z)\cong\Z^{4,4}$ and is given with respect to the
basis $e_1\wedge e_2\wedge e_3 \wedge e_4$,
$e_i\wedge e_j, 1$. 
Note that in \cite{nawe00} we have used a normalization
of the scalar product on $H^2(T_0,\Z)$ which differs by a factor of $V$
from the above.

For later use we note that analogously to (\ref{spinor}) one
determines
\begin{equation}\label{subspinor}
s(C) =
\left( \begin{array}{crc} 1  & 0  & 0\\
                          -\check c & \id & 0\\
                          -{\|C\|^2\over2}  & c  & 1 \end{array}\right)
\end{equation}
as the triality conjugate of
$$
v(C) =
\left( \begin{array}{rc} \id & 0\\
                          -C  & \id \end{array}\right)
\in O^+(4,4),
$$
where $c$ is the row vector with components $c_{ij}=C_{ij}$
with respect to $e_i\wedge e_j$.

The sigma model on $T_0$ with $B=0$ is described by the lattice
$L(T_0)$ and the positive definite four-plane $x_0\subset
H^{even}(T_0,\R)$ which is left invariant by the Hodge star
operator $\ast$. The latter is given by
\begin{equation}\label{fourplane}
x_0=\span_\R\left\{
\begin{array}{ll}
1 +  e_1\wedge e_2\wedge e_3\wedge e_4,\quad
&  e_1\wedge e_3 + e_4\wedge e_2,\\
e_1\wedge e_2 + e_3\wedge e_4 ,
&e_1\wedge e_4 + e_2\wedge e_3 
\end{array}\right\}.
\end{equation}
For arbitrary sigma model parameters $(\Lambda,B)$, $x_0$ as in 
(\ref{fourplane}) remains the $+1$ eigenspace of $\ast$, but   we
have $H^{even}(T,\Z)\cong s(\Lambda,B)L(T_0)=:L$.
A point in the cover $\widetilde{\mathcal M}$ of the moduli space
is described by the relative position of $L$ with respect to $x_0$, i.e.\ the
pair $x_0,s(\Lambda,B)$. 
Since only the relative position
counts, we can use $Rx_0$ and $Rs(\Lambda,B)$ with arbitrary
$R\in O^+(4,4)$. To avoid confusion one
should note that, as mentioned above, in  \cite{nawe00} we have used
$Rs(\Lambda,B)=s(V^{-1/4}\Lambda,B)$.  
For the description of mirror symmetry on orthogonal tori, 
however, it is more convenient
to choose $R=\id$. 

Let us now determine the lattice automorphism that acts as mirror
symmetry. It suffices to consider $B=0$ and a torus $T$ with defining matrix
$\Lambda=\diag(r_1,\dots,r_4), r_i>0$. The generators of its even 
cohomology group $H^{even}(T,\Z)$ will be denoted
$\upsilon=\mu_1\wedge\dots\wedge\mu_4$,
$\mu_i\wedge\mu_j$, $\upsilon^0=1$.
By the above we need to find a map that leaves $H^{even}(T,\Z)$
and the four--plane (\ref{fourplane}) invariant and induces 
$r_1\mapsto (r_1)^{-1}, r_2\mapsto (r_2)^{-1}$
on the torus parameters.   
To this end, substituting
$e_i=r_i\mu_i$   into
(\ref{fourplane}) one in particular finds  
\begin{eqnarray*} 
\gamma_{MS}(T_0) : \left\{
\begin{array}{lclrcl}
\pm\,\upsilon^0 & \longleftrightarrow& \pm\,\mu_1\wedge \mu_2,\quad
&\pm\,\mu_1\wedge \mu_3 & \longleftrightarrow& \pm\,\mu_2\wedge \mu_3,\\[3pt]
\pm\,\upsilon & \longleftrightarrow& \pm\,\mu_3\wedge \mu_4,
&\pm\,\mu_4\wedge \mu_2 & \longleftrightarrow& \pm\,\mu_1\wedge \mu_4 
\end{array}
\right.\hspace*{-3em}
\end{eqnarray*}
for the base point
of the moduli space given by $T_0$ and $B=0$ \cite{na00b}.
To fix the signs, recall   that   
T-duality in the $x_1,x_2$ fiber of our $T^2$ fibration of $T$, 
as automorphism of the Grassmannian of four-planes in
$H^1(T_0,\R)\oplus H_1(T_0,\R)$, 
acts by
conjugation with the element 
$\sigma = \left(\diag(-1,-1,1,1),\id\right) \in
SO(4)_{left}\times SO(4)_{right}\subset O^+(4,4)$. 
Correspondingly, its action on $H^{even}(T_0,\R)$ is given
by the spinor representation $s(\sigma)$ of this group element. 
Since $\sigma$ is a rotation by $\pi$, the square of $s(\sigma)$
is $-\id$. We will argue that (up to an irrelevant overall sign) 
\begin{eqnarray}\label{torms}
\gamma_{MS}(T_0) : \left\{
\begin{array}{rcrcl} 
\upsilon^0& \longmapsto &\mu_1\wedge \mu_2&\longmapsto & -\upsilon^0,\\[2pt]
\upsilon& \longmapsto &\mu_3\wedge \mu_4 & \longmapsto & -\upsilon ,\\[2pt] 
\mu_1\wedge \mu_3 & \longmapsto & \mu_2\wedge \mu_3&\longmapsto 
&-\mu_1\wedge \mu_3,\\[2pt]
\mu_4\wedge \mu_2 & \longmapsto& \mu_1\wedge \mu_4& \longmapsto&
-\mu_4\wedge \mu_2,
\end{array}
\right.\hspace*{-3em}
\end{eqnarray}
where the lower two lines of (\ref{torms}) are induced by
$\mu_1\mapsto\mu_2\mapsto-\mu_1$. This can be explained as follows.
Above, we have described mirror symmetry by the sign change of
two left handed current components $(j_1,j_2)\mapsto(-j_1,-j_2)$,
whereas the right handed components are unchanged. Since only
the relative rotation by $\pi$ between the two chiralities
is important, one may as well consider the maps
$(j_1,j_2)\mapsto(j_2,-j_1)$ for the left handed components
and $(\bar\jmath_1,\bar\jmath_2)\mapsto(-\bar\jmath_2,\bar\jmath_1)$ for the
right handed ones. By the above, this rotation is described by
$$
v_{ms}=\exp(\pi\Omega/2), \quad\mbox{ where } \quad
\Omega =
\left( \begin{array}{cc} 0 & \Omega_{12}\\
                          \Omega_{12}  & 0 \end{array}\right)
\in o^+(4,4),
\;
\Omega_{12} =
\left( \begin{array}{rr} 0 & -1\\
                      1 & 0 \end{array}\right).
$$
We use (\ref{spinor}) and (\ref{subspinor})
to translate this into the half spinor representation and find
$$
s_{ms} = \exp(\pi\widetilde\Omega_{12}/2), \quad\mbox{ where } \quad
\widetilde\Omega_{12} =
\left( \begin{array}{ccc} 0 & \;-\check\omega_{12}^T\; & 0\\
                      \check\omega_{12} & 0 & \omega_{12}\\
                          0 & -\omega_{12}^T  & 0 \end{array}\right),
$$
and  $\omega_{12}, \check\omega_{12}$ are obtained from $\Omega_{12}$
as explained above for $b,\check b, B$.
Since this gives the transformation with respect to the basis
$\upsilon, \mu_i\wedge\mu_j,\upsilon^0$, the
mirror symmetry transformation given by $s_{ms}$ can be
written as
\begin{eqnarray}\label{tormsz3}
\gamma_{MS}(T_0^\prime) : \left\{
\begin{array}{rcrrcl}
\upsilon^0& \longmapsto &\mu_1\wedge \mu_2,
&\quad\mu_1\wedge \mu_2& \longmapsto & -\upsilon^0 ,\\[2pt]
\upsilon& \longmapsto &\mu_3\wedge \mu_4,
&\mu_3\wedge \mu_4 & \longmapsto & -\upsilon  ,\\[2pt]
\mu_1\wedge \mu_3 & \longmapsto & \mu_1\wedge \mu_3, &
\mu_2\wedge \mu_3&\longmapsto & \mu_2\wedge \mu_3 ,\\[2pt]
\mu_4\wedge \mu_2 & \longmapsto& \mu_4\wedge \mu_2, &
\mu_1\wedge \mu_4& \longmapsto& \mu_1\wedge \mu_4.
\end{array}
\right.\hspace*{-3em}
\end{eqnarray}
To obtain the   mirror map $\gamma_{MS}(T_0)$ corresponding
to the original transformation $(j_1,j_2)\mapsto(-j_1,-j_2)$,
$(\bar\jmath_1,\bar\jmath_2)\mapsto(\bar\jmath_1,\bar\jmath_2)$ one composes
$\gamma_{MS}(T_0^\prime)$ with the classical symmetry
$(j_1,j_2)\mapsto(j_2,-j_1)$, 
$(\bar\jmath_1,\bar\jmath_2)\mapsto(\bar\jmath_2,-\bar\jmath_1)$ and finds
(\ref{torms}).

When we discuss orbifolds with respect to threefold rotations in fiber and
base, the transformation (\ref{torms}) is not always applicable.
For such rotations, the forms $\upsilon, \upsilon^0$,
$\mu_1\wedge \mu_2$, and $\mu_3\wedge \mu_4$ are invariant, but not
necessarily the others. In particular, the four plane (\ref{fourplane})
is not invariant. The transformation $\gamma_{MS}(T_0^\prime)$
is well behaved in all cases, however.
When we work with orbifolds of the corresponding fibered tori we always
keep $\mu_1,\mu_2$ as generators of periods in the fiber and
$\mu_3,\mu_4$ as generators of periods in the base,
such that $\gamma_{MS}(T_0^\prime)$ lifts to a symmetry of the conformal
field theory which commutes with the symmetries used for orbifolding.

We stress that $\gamma_{MS}(T_0), \gamma_{MS}(T_0^\prime)$ are 
lattice automorphisms
of order $4$, and $\gamma_{MS}(T_0)\circ\gamma_{MS}(T_0)=-\id$.   
It has also been observed before that $\gamma_{MS}(T_0)$  can be
understood as hyperk\a hler rotation in $\widetilde{\mathcal M}$ 
\cite{bbrm98,di99,brsa98,achy00,achy01}, in full agreement with the above.

With respect to the complex structure $\mathcal I$ given by
$(e_1-ie_3)\wedge(e_2+ie_4)$, the second line in
(\ref{fourplane})
corresponds to the complex structure and the first   to its orthogonal
complement. $\gamma_{MS}(T_0)$ exchanges the two, in accord with the
general notion of mirror symmetry discussed in the Introduction.
Moreover, the $T^2$ fibration with fiber coordinates
$x_1,x_2$
is elliptic with respect to the complex structure $\mathcal J$ given by
$(e_1+ie_2)\wedge (e_3+ie_4)$ (with a holomorphic section),  and therefore
it is special Lagrangian with respect to $\mathcal I$ (c.f. \cite{hala82}).
Hence (\ref{torms}) also describes a version of 
mirror symmetry in the sense of \cite{syz96,mo96}, obtained from
$r_1\mapsto (r_1)^{-1}, r_2\mapsto (r_2)^{-1}$, as in \cite{vawi95}.
\section{The Mirror Map for Kummer Surfaces}\label{kummer}
Recall the classical Kummer construction of $K3$: Given a four-torus
$T$, we have a $\Z_2$ symmetry induced by multiplication with $-1$ on
$\R^4$. By minimally resolving the $16$ singularities of the 
corresponding $\Z_2$
orbifold of $T$ and assigning volume zero to all exceptional divisors
in the blow up we obtain an orbifold limit of $K3$, a Kummer surface $X$.
In particular, there is a rational map $\pi:T\longrightarrow X$ of 
degree $2$ which is defined outside the fixed points.
The $T^2$ fibration of $T$ used in Sect.~\ref{tori} induces a
$T^2$ fibration $p:X\longrightarrow\P^1$ which is elliptic with respect
to $\pi_\ast \mathcal J$ and therefore special Lagrangian with respect
to $\pi_\ast \mathcal I$. 
Note that the holomorphic section is not the
$\pi_\ast$ image of the section in our fibration of $T$. 
We rather have to make sure that the fibration can be written in the 
Weierstra\ss\ form, such that each singular
fiber can be labeled by its Kodaira type \cite{ko63b}. The Poincar\'e
duals of the generic fiber and generic section are given in
(\ref{elliptic}) below. Apart 
from the   behavior at the four 
singular fibers, mirror symmetry as discussed above
is induced by mirror symmetry on the torus. This was in fact proven more
generally in \cite{vawi95} and generalized to the Strominger/Yau/Zaslow
conjecture in \cite{syz96}.

It again suffices to specialize to the standard torus $T_0$, consider
the corresponding Kummer surface  $X_0$, and
determine the   automorphism of the
lattice $L(X_0)$ that acts as mirror symmetry at this base point.
To this end, let us recall the description of $L(X_0)$ as found in
\cite{ni75,nawe00}.  
The orbifolding map $\pi$ induces an injective map $\pi_\ast$ on
cohomology such that\footnote{Given a
lattice $\Gamma$, by $\Gamma(n)$ we denote the same $\Z$ module as
$\Gamma$   with quadratic form scaled by a factor of  $n$.}
$\pi_\ast H^{even}(T_0,\Z)$ $\cong H^{even}(T_0,\Z)(2)$.
We embed $H^{even}(T_0,\Z)(2)$ in $H^{even}(X_0,\Z)$ by rescaling
$L(T_0)=H^{even}(T_0,\Z)$
with $\sqrt{2}$. With this convention we have
$H^{even}(X_0,\R)=\pi_\ast H^{even}(T_0,\R)\perp\span_\R\{ E_i\vert i\in I\}$,
where $I$ labels the 16 fixed points of the $\Z_2$ orbifolding,  
and the $E_i$ project to the Poincar\'e duals of the exceptional divisors
in the blow up of these fixed points (see below). Since each divisor is a
rational curve of self-intersection number $-2$ on $X_0$, the $E_i$
generate a lattice $\Z^{16}(-2)\subset H^{even}(X_0,\R)$.
On $I$ one
finds an affine $\F_2^4$ geometry\footnote{As usual, $\F_p$, $p$ prime, 
denotes the
unique
finite field with $p$ elements.} \cite{ni75}, which we use to
label the fixed points.
The four--plane $x_0\subset H^{even}(T_0,\R)$ given in (\ref{fourplane})
remains unchanged.

By $\Pi$ we denote the so-called \textsl{Kummer lattice}
$$
\vphantom{\sum_{i\in H}}
\Pi
:=\span_\Z\left\{ E_i, i\in I; \quad
{\textstyle{1\over2}} \smash{\sum_{i\in H}}
E_i, H\subset I \makebox{ a hyperplane }
\right\}.
$$
Its projection $\widehat\Pi\cong\Pi$ 
onto $H^2(X_0,\Z)$ is the minimal primitive
sublattice   which contains all Poincar\'e duals $\widehat E_i, i\in I,$ 
of exceptional divisors.

Let $P_{j,k}:=\span_{\F_2}(f_j,f_k)\subset \F_2^4$ with $f_j\in \F_2^4$
the $j^{th}$ standard basis vector and  $Q_{j,k} := P_{l,m}$ such that
$\{j,k,l,m\}=\{1,2,3,4\}$.
Then\footnote{We remark that in \cite{nawe00} we missed to 
exchange $P_{j,k}$ with $Q_{j,k}$, which amounts to translation
from homology to cohomology by Poincar\'e duality.} \cite{ni75}
$$
M :=
\left\{ {\textstyle{1\over\sqrt2}}\mu_j\wedge \mu_k
- {\textstyle{1\over2}}\smash{\sum_{i\in Q_{j,k}}} E_{i+l}, l\in I
\right\}\mbox{ and } \Pi
\vphantom{\sum_{i\in P_{j,k}}}
$$
generate a lattice isomorphic to $H^{2}(X_0,\Z)$.

The lattice $L(T_0)(2)=H^{even}(T_0,\Z)(2)$ 
and $M$ belong to $L(X_0)$, but the
$E_i$ do not, since $\pi_\ast L(T_0)\perp\Pi$ cannot be embedded as
sublattice into $L(X_0)$.
Instead,  
$L(X_0)=\span_\Z\{\widehat{M}\cup\Pi_0\}$,
where
\begin{eqnarray}\label{hat}
\widehat{M}
&:=& M \cup
\left\{ 
\widehat\upsilon^0
:={\textstyle{1\over\sqrt2}}\upsilon^0 + {\textstyle{1\over
4}}\smash{\sum_{i\in I}E_i}+\sqrt2\upsilon;\quad
\widehat{E}_i:= E_i+{\textstyle{1\over \sqrt2}}\upsilon  , i\in I
\right\}\nonumber \\[2mm]
\mbox{and }\quad\Pi_0
&:=&\left\{\pi\in\Pi \mid \forall\; m\in\widehat{M}:\,
\langle\pi,m\rangle\in\Z \right\} 
\end{eqnarray}
\cite{nawe00}. It is important to note that $\widehat E_i$ is the two-form
contribution to $E_i$, and vice versa $E_i$ is the orthogonal projection 
of the lattice vector $\widehat E_i$ onto  $(\pi_\ast H^{even}(T_0,\R))^\perp$.
The observation that this
gives the unique consistent embedding of $\pi_\ast L(T_0)$
into $L(X_0)$ implies that the
B-field in a $\Z_2$ orbifold CFT on $K3$ has
value $1/2$ in direction of each exceptional divisor of the blow up
\cite{as95,nawe00}. This observation generalizes to all  orbifold
CFTs on $K3$ \cite{we00}.
The lattice of two-form contributions to vectors in $\Pi$ is denoted
$\widehat\Pi$, in the following.

With the above description of $H^{even}(X_0,\Z)$ one checks that the 
Poincar\'e duals of generic fiber and generic
holomorphic section in our elliptic
fibration $p:X\longrightarrow\P^1$ are given by
\begin{eqnarray}\label{elliptic}
&&\sqrt2 \mu_3\wedge\mu_4,\quad \\
&&\inv{\sqrt2} \mu_1\wedge\mu_2 
- \inv2 \left(   E_{(0,0,0,0)} +   E_{(0,0,1,0)} + 
 E_{(0,0,0,1)} +   E_{(0,0,1,1)} \right), \nonumber
\end{eqnarray}
respectively.

We will now determine the induced mirror map 
$\gamma_{MS}(X_0)\in$ $\Gamma=$  $Aut(L(X_0))$ $\cong O^+(4,20;\Z)$.
The geometric description of mirror symmetry implies that the
action on $L(T_0)(2)$ is induced by the action of $\gamma_{MS}(T_0)$
on $L(T_0)$. To extend it to all of $L(X_0)$ we have to find
images of $E_{i}, i\in I,$ in $\Pi\otimes\Q$, such that
the induced linear map is an automorphism of $\Gamma$.
We claim that with arbitrary $K_0,M_0\in\F_2$ and $t_0:=(0,0,K_0,M_0)\in I$ 
the following map will do:
\begin{equation}\label{k3ms}
\begin{array}{l}
\forall\;(I,J,K,M)\in\F_2^4:\\[3pt]
\ds\quad\quad\quad
\gamma_{MS}(X_0)(E_{(I,J,K,M)}) :=
{\textstyle{1\over2}} \sum_{i,j\in\F_2} (-1)^{iI+jJ}
 E_{(i,j,K,M)+t_0}.
\end{array} 
\end{equation}
First, it is easy to see that this map preserves scalar products and 
acts as  involution on $\Pi_0$.
Since $\Pi_0$ is generated by 
$E_{i}\pm E_j, i,j\in I,$ and ${\textstyle{1\over2}} \sum_{i\in H} E_{i}, H\subset I$
a hyperplane,
one finds  that (\ref{k3ms}) maps
$\Pi_0$ into itself. Next, we check that
$M\subset\widehat{M}$ is mapped into $H^{even}(X_0,\Z)$. Namely,  
there are $\pi_{a,b}\in\Pi_0$ such that
for all $I,J,K,L\in\F_2$
\begin{eqnarray*}
\noalign{$\displaystyle \gamma_{MS}(X_0)
\left( {\textstyle{1\over\sqrt2}} \mu_1\wedge \mu_3
- {\textstyle{1\over2}} 
\smash{\sum_{i\in Q_{1,3} }} E_{i+(I,J,K,M)} \right) $}\\
&=&  {\textstyle{1\over\sqrt2}} \mu_2\wedge \mu_3 
- {\textstyle{1\over2}}  \sum_{i,m\in\F_2}(-1)^{iI} E_{(i,0,K,m)+t_0}  \\
&=& {\textstyle{1\over\sqrt2}} \mu_2\wedge \mu_3 
-{\textstyle{1\over2}} \sum_{i\in Q_{2,3}} E_{i+(0,0,K,0)+t_0} + I\pi_{1,3},\\
\noalign{$\displaystyle 
\gamma_{MS}(X_0)\left( {\textstyle{1\over\sqrt2}} \mu_4\wedge \mu_2
- {\textstyle{1\over2}} 
\smash{\sum_{i\in Q_{2,4} }} E_{i+(I,J,K,M)} \right) $}\\
&=& {\textstyle{1\over\sqrt2}} \mu_1\wedge \mu_4 
- {\textstyle{1\over2}}  \sum_{j,k\in\F_2} (-1)^{jJ} E_{(0,j,k,M)+t_0}  \\
&=& {\textstyle{1\over\sqrt2}} \mu_1\wedge \mu_4
- {\textstyle{1\over2}} \sum_{i\in Q_{1,4} } E_{i+(0,0,0,M)+t_0} + J\pi_{2,4},
\end{eqnarray*}
and $\pi_{(I,J,K,M)} \in\Pi_0$ with
\begin{eqnarray*}
\noalign{$\displaystyle 
\gamma_{MS}(X_0)\left( {\textstyle{1\over\sqrt2}} \mu_3\wedge \mu_4
- {\textstyle{1\over2}} \smash{\sum_{i\in Q_{3,4} }} E_{i+(I,J,K,M)}
\right) $}\\
\hphantom{\sum}
&=& -{\textstyle{1\over\sqrt2}} \upsilon  - E_{(0,0,K,M)+t_0} 
\;\stackrel{(\ref{hat})}{=}\;  -\widehat{E}_{(0,0,K,M)+t_0},\\
\noalign{$\displaystyle 
\gamma_{MS}(X_0)\left( {\textstyle{1\over\sqrt2}} \mu_1\wedge \mu_2
- {\textstyle{1\over2}} \smash{\sum_{i\in Q_{1,2} }} 
E_{i+(I,J,K,M)} \right) $}\\
&=& -{\textstyle{1\over\sqrt2}} \upsilon^0  
- {\textstyle{1\over4}}  \sum_{k,m\in\F_2} 
\left( E_{(0,0,k,m)+t_0} + (-1)^I E_{(1,0,k,m)+t_0} \right.\\
&&\left.
\hphantom{{\textstyle{1\over\sqrt2}} \upsilon^0 }
      + (-1)^J E_{(0,1,k,m)+t_0} + (-1)^{I+J} E_{(1,1,k,m)+t_0} \right)\\
&=& -{\textstyle{1\over\sqrt2}} \upsilon^0  
- {\textstyle{1\over4}}  \sum_{i\in I} E_{i} + \pi_{(I,J,K,M)} .
\end{eqnarray*}
Since $\gamma_{MS}(X_0)_{\mid\Pi_0}$ 
is an involution, and 
$\gamma_{MS}(X_0)\circ\gamma_{MS}(X_0)_{\mid\Pi_0^\perp}
= -\id$,
this suffices to prove consistency. 
From the above it also follows that up to automorphisms of
$\left(\pi_\ast H^{even}(T_0,\Z)\right)^\perp\cap H^{even}(X_0,\Z)$,
(\ref{k3ms}) gives the only consistent
maps $\gamma_{MS}(X_0)$. We will be more precise about this point at the end 
of  Sects.~\ref{kummertype} and \ref{mckay}.

Let us consider the actual geometric action of $\gamma_{MS}(X_0)$. From the
above  
we can easily write out the map on 
$H^{even}(X_0,\Z)=\span_\Z\{\widehat M\cup\Pi_0\}$. 
Hence we have in particular
found an explicit continuation of mirror symmetry as 
induced by fiberwise T--duality to the four singular fibers of 
$p:X\longrightarrow\P^1$ over $x_3\in\{0,r_3/2\}, x_4\in\{0,r_4/2\}$, i.e.\
with labels $(K,M)\in\F_2^2$.
Each of these singular fibers is of type $I_0^\ast$ in Kodaira's classification
\cite[Th.6.2]{ko63b} with components dual to
$$
\widehat E_{(i,j,K,M)}, i,j\in\F_2, \mbox{ and }
C_{K,M}:={\textstyle{1\over\sqrt2}} \mu_3\wedge \mu_4
- {\textstyle{1\over2}}\sum_{i\in Q_{3,4}} \widehat E_{i+(0,0,K,M)}.
$$ 
The latter form $C_{K,M}$ corresponds to the center node of the 
$\widehat D_4$ type Dynkin diagram describing  $I_0^\ast$. 
For simplicity, let us set $t_0=(0,0,0,0)$.
Since for
suitable $\pi_{(I,J,K,M)}^\prime \in\Pi_0$ and with the generator
$\widehat\upsilon=\sqrt2\upsilon$   of $H^4(X_0,\Z)$ (c.f.
\cite{nawe00}),  
\begin{eqnarray*}
\gamma_{MS}(X_0)(  \widehat{E}_{(I,J,K,M)}) 
&\stackrel{(\ref{hat})}{=}&
\gamma_{MS}(X_0)\left(  
{\textstyle{1\over\sqrt2}} \upsilon+E_{(I,J,K,M)} 
\right)\\
&=& {\textstyle{1\over\sqrt2}} \mu_3\wedge \mu_4
+ {\textstyle{1\over2}} \left(  E_{(0,0,K,M)} + (-1)^I E_{(1,0,K,M)} 
\right.  \\
&&\left.      + (-1)^J E_{(0,1,K,M)} 
+ (-1)^{I+J} E_{(1,1,K,M)} \right)  \\
&=&    C_{K,M} + \pi_{(I,J,K,M)}^\prime 
- \delta_{I,0}\delta_{J,0}\widehat\upsilon, \\
\gamma_{MS}(X_0)(C_{K,M})&=&  
-\widehat E_{(0,0,K,M)}  - \sqrt2 \mu_3\wedge \mu_4,
\end{eqnarray*}
we see that up to signs and possible corrections in $\Pi_0$ and
$\pi_\ast H^{even}(T_0,\Z)$, $\gamma_{MS}(X_0)$ exchanges the center node of 
$I_0^\ast\cong \widehat D_4$ with each of its four legs.

Note also that for $t_0=(0,0,0,0)$ the Poincar\'e duals 
(\ref{elliptic}) of generic fiber and generic section of our elliptic fibration
are simply mapped onto 
$-\widehat\upsilon, -\widehat\upsilon^0+\widehat\upsilon$
under $\gamma_{MS}(X_0)$, where $\widehat\upsilon,\widehat\upsilon^0$
are the generators of $H^4(X_0,\Z), H^0(X_0,\Z)$,
respectively (see (\ref{hat}) and \cite{nawe00}).

Next, let us investigate the monodromy $[m], m\in SL(2,\Z)$, of the regular 
two-tori in our fibration $p:X\longrightarrow\P^1$ around a singular fiber,
where $[m]$ denotes the conjugacy class of $m$ in $SL(2,\Z)$.
Since $\gamma_{MS}(X_0)$ acts by $S\in SL(2,\Z)$
on the modular parameter of the fiber,
$$
S=\left( \begin{array}{rr}0&-1\\1&0 \end{array}\right),
$$
the monodromy should transform as
\begin{equation}\label{mimo}
m\quad\longmapsto\quad \left(m^T\right)^{-1} \quad = \quad SmS^{-1} ,
\end{equation}
i.e.\ $[m]$ should remain invariant 
under mirror symmetry. In the present case, all singular fibers are of
type $I_0^\ast$, hence by \cite[Th.9.1]{ko63b} their monodromy is 
\begin{equation}\label{kumo}
m =\left( \begin{array}{cc}-1&0\\0&-1 \end{array}\right),
\end{equation}
which is even invariant under (\ref{mimo}). This is in accord with our 
construction, since our geometric interpretation of
a $K3$ theory obtained as $\Z_2$ orbifold
on a Kummer surface with $B=0$ for the underlying toroidal theory
is indeed
mapped into another such geometric interpretation. The mirror $K3$ is hence
expected to be a singular Kummer surface again, with singular fibers of
type $I_0^\ast$, and with monodromy (\ref{kumo}) around each of them.
\section{The Mirror Map for Kummer Type Surfaces}\label{kummertype}
In this section, we give the action of mirror symmetry on    
Kummer type $K3$ surfaces obtained as orbifold limits 
$\widetilde{T/\Z_N}, N\in\{3,4,6\}$ of $K3$. 
Since the proofs are analogous to the
one for $N=2$ that has been
discussed at length in the previous section,
we restrict ourselves to a presentation of the results. For
explicit proofs, one needs the description
of $L(X_0)$ in terms of $\pi_\ast L(T_0)^{\Z_N}\cong L(T_0)^{\Z_N}(N)$  and the
exceptional divisors obtained by minimally resolving 
all singularities, as given
in \cite{nawe00,we00}.
 
Recall the $\Z_N$ orbifold construction of $K3$, $N\in\{3,4,6\}$. Let
$\zeta_N$ denote a generator of $\Z_N$, where $\Z_N$ is realized as
group of $N^{th}$ roots of unity in $\C$, 
and   $z_1,z_2$   complex 
coordinates on $T$, such that $T=\widetilde T^2\times T^2$ with elliptic
curves $\widetilde T^2, T^2$. If both curves are $\Z_N$ symmetric
(note that the metric in general 
need not be diagonal with respect to $z_1,z_2$),
there is an algebraic $\Z_N$ action on $T$ given by
$$
\Z_N\ni\zeta_N^l:\quad \zeta_N^l.(z_1,z_2)=(\zeta_N^l z_1,\zeta_N^{-l}z_2).
$$
By minimally resolving 
the singularities of $T/\Z_N$ and assigning volume zero
to all components of the
exceptional divisors we obtain a $\Z_N$ orbifold limit $X$ of $K3$.
The fixed point set of $\Z_N$ on $T$ will be denoted $I$ in general, and
$n(t)$ is the order of the stabilizer group of $t\in I$.
For $N=3$ we have $I\cong\F_3^2$, and since the $\Z_4,\Z_6$
orbifold limits can be obtained from the $\Z_2,\Z_3$ orbifold limits
by modding out an algebraic $\Z_2$ action,
for $N=4,6$ we use
$I = \F_2^4/\sim$ and $I = \F_2^4\cup\F_3^2/\sim$, respectively, where 
$\sim$ denotes the necessary identifications.
To assure compatibility of the $\Z_N$ action with our fibration
$p:T\longrightarrow T^2$, we choose
$z_1=x_1+ix_2, z_2=x_3+ix_4$ and obtain an induced fibration 
$p:X\longrightarrow \P^1$ as in the $\Z_2$ case. 
We can also restrict considerations
to appropriate standard tori $T_0, T_0^\prime$
for $N=4, N\in\{3,6\}$, respectively, 
$X_0^{(\prime)}=\widetilde{T_0^{(\prime)}/\Z_N}$.
Here, $T_0=\R^4/\span_\Z\{e_1,\dots,e_4\}$ with  an orthonormal basis 
$e_1,\dots,e_4$ as introduced 
in Sect.~\ref{tori},   and 
$$
T_0^\prime=\R^4/\span_\Z\{\lambda_1,\dots,\lambda_4\}
\quad\mbox{ with }\quad\left\{
\begin{array}{ll}
\mu_1=e_1, \quad
& \mu_2=\inv2 e_1+\inv[\sqrt3]{2}e_2,\\[2pt]
\mu_3 = e_3, 
&\mu_4 = \inv2 e_3+\inv[\sqrt3]{2}e_4
\end{array}\right.
$$
for the dual basis $\mu_1,\dots,\mu_4$.
The $\Z_3$ action then is given by
$$
\zeta_3:\left\{\begin{array}{lcllcl}
\mu_1&\longmapsto& \mu_2-\mu_1, \quad&\mu_2&\longmapsto&-\mu_1,\\[2pt]
\mu_3&\longmapsto&-\mu_4,&\mu_4&\longmapsto& \mu_3-\mu_4.
\end{array}\right.
$$
The exceptional divisor in the minimal resolution of a $\Z_{n(t)}$ type 
fixed point $t\in I$ has $n(t)-1$ irreducible
components with intersection matrix the negative of
the Cartan matrix of the Lie group $A_{n(t)-1}$. The projections to 
$(\pi_\ast L(T_0)^{\Z_N})^\perp$  of the
Poincar\'e duals of these $(-2)$ curves
are denoted $E_t^{(l)}, l\in\{1,\dots,n(t)-1\}$, such that
$$
\left\langle E_t^{(l)},E_t^{(m)}\right\rangle = \left\{
\begin{array}{cl} 
-2&\mbox{ for } l=m,\\
1&\mbox{ for } |l-m|=1,\\
0&\mbox{ otherwise.}
\end{array}\right.
$$
We define 
$$
E_t:=\sum_{l=1}^{n(t)-1}lE_t^{(l)},\quad
E_t^{(0)}:=-\!\sum_{l=1}^{n(t)-1}E_t^{(l)},
$$
as well as a $\Z_{n(t)}$ action on 
$\E_t:=\span_\Z\{ E_t^{(l)}, l\in\{1,\dots,n(t)-1\}\}$
generated by
\begin{equation}\label{thetadef}
\vartheta(E_t^{(l)}) := \left\{ 
\begin{array}{ll}
E_t^{(l+1)} & \mbox{ if } l<n(t)-1,\\[2pt]
E_t^{(0)}  & \mbox{ if } l=n(t)-1,
\end{array}\right.
\mbox{ such that } (\vartheta-1)E_t=n(t)E_t^{(0)}.
\end{equation}
As in Sect.~\ref{kummer} it suffices to determine the image of each
$E_t^{(l)}$ under the extension of the mirror map of the underlying 
toroidal theory. For the $\Z_4$ orbifold, the latter is  
$\gamma_{MS}(T_0)$ as in (\ref{torms}), and for the $\Z_3$ and $\Z_6$
orbifolds we have to use $\gamma_{MS}(T_0^\prime)$ as defined in
(\ref{tormsz3}).

We list the singular fibers of $p:X\longrightarrow\P^1$
in terms of Kodaira's classification \cite[Th.6.2]{ko63b}, 
all of which are non-stable. Recall also
the labels of the corresponding extended Coxeter Dynkin diagrams:
\begin{eqnarray*}
\Z_3:\;  IV^\ast+IV^\ast+IV^\ast, \quad 
\Z_4:\; I_0^\ast +  III^\ast+  III^\ast,
\quad \Z_6:\; I_0^\ast+II^\ast+IV^\ast;\\
IV^\ast\cong \widehat E_6,\quad  I_0^\ast\cong\widehat D_4, \quad
III^\ast \cong \widehat E_7, \quad  II^\ast \cong \widehat E_8.
\end{eqnarray*}
By construction, our map $\gamma_{MS}(X_0^{(\prime)})$ 
acts fiberwise (c.f. (\ref{k3ms})), 
hence it suffices to 
specify the map on each type of singular fibers. Accordingly,  
fixed points are only labeled by the fiber coordinates $x_1,x_2$ in the 
following. 

Fibers of type $I_0^\ast$ have been discussed in Sect.~\ref{kummer}.
Type
$IV^\ast$ occurs in both $\Z_3$ and $\Z_6$ orbifolds  and contains three 
$A_2$ type exceptional divisors with components Poincar\'e dual to
$E_t^{(l)}, l\in\{1,2\}, t\in\F_3$. We find
\begin{equation}\label{msz3}
t\in\F_3, l\in\{1,2\}:\quad
\gamma_{MS}(X_0^\prime)(E_t^{(l)}) 
= -\inv{3} \sum_{k\in\F_3} 
\vartheta^{l+kt} (E_k),
\end{equation}
where we have chosen an origin $0\in\F_3$ and the standard scalar
product on $\F_3$.

For $III^\ast$, which occurs in the $\Z_4$ orbifold, we have two $A_3$
type exceptional divisors giving $E_i^{(l)}, l\in\{1,2,3\}, 
i\in\{(0,0),(1,1)\}$, and one $A_1$ type exceptional divisor corresponding
to $E_{(1,0)}$. Then
\begin{eqnarray}
i\in\F_2, l\in\{1,2,3\}:\nonumber\\
\gamma_{MS}(X_0)(E_{(i,i)}^{(l)}) 
&=& -\inv{2} (-1)^{l+i} (E_{(1,0)})
- \inv{4} \sum_{k\in\F_2} \vartheta^{l+2ik} (E_{(k,k)}),\label{msz4}\\
\gamma_{MS}(X_0)(E_{(1,0)}) 
&=&  
-\inv{4}(\vartheta+\vartheta^3) 
\sum_{k\in\F_2} \vartheta^{k} (E_{(k,k)}).\nonumber
\end{eqnarray}
Finally, for $II^\ast$ we have one $A_5,A_2,A_1$ type exceptional divisor
each, corresponding to $E_0^{(l)}, l\in\{1,\dots,5\}, 
E_1^{(l)}, l\in\{1,2\}, E_{(1,0)}$. Here,
\begin{eqnarray}
l\in\{1,\dots,5\}:\quad
\gamma_{MS}(X_0^\prime)(E_0^{(l)}) 
&=& -\inv{2} (-1)^{l} (E_{(1,0)}) - \inv{3}  \vartheta^{l} (E_1)
- \inv{6}   \vartheta^{l} (E_0),\nonumber\\
l\in\{1,2\}:\quad
\gamma_{MS}(X_0^\prime)(E_1^{(l)}) 
&=&   \inv{3}  \vartheta^{l} (E_1)
- \inv{6} (\vartheta^0+\vartheta^3)  \vartheta^{l} (E_0),\nonumber\\
\gamma_{MS}(X_0^\prime)(E_{(1,0)}) 
&=&   - \inv{2}   E_{(1,0)} 
- \inv{6} (\vartheta+\vartheta^3+\vartheta^5)  (E_0).\label{msz6}
\end{eqnarray}
To prove the above,   one   
has to check that  scalar
products are preserved and that
the generator $\widehat \upsilon^0$ of $H^0(X_0^{(\prime)},\Z)$ 
(see (\ref{upsilon0}))
is mapped onto a lattice vector
as well as
\begin{equation}\label{ehat}
\forall\,t\in I, l\in\{1,\dots,n(t)-1\}:\quad
\widehat E_t^{(l)}= E_t^{(l)}+\inv{n(t)}\widehat\upsilon,
\quad \widehat\upsilon=\sqrt N \upsilon
\end{equation}
(recall $E_t^{(l)}\not\in L(X^{(\prime)}_0)$, in general, and that 
$\widehat\upsilon=\sqrt N \upsilon$ generates $H^4(X^{(\prime)}_0,\Z)$; see
\cite{nawe00,we00} and compare to (\ref{hat})).
This is a 
straightforward calculation using \cite{we00}.

For later convenience, (\ref{k3ms}), (\ref{msz3})-(\ref{msz6})  can
be  summarized in the following formula:
\begin{eqnarray}\label{generalms}
&&t\in I \mbox{ a fixed point of type }\Z_{n(t)}:\quad  \\
&&\hphantom{fixed point }
\gamma_{MS}(X_0^{(\prime)})(E_t^{(L)}) 
= -\inv{N}\sum_{\stackrel{\scriptstyle k\in \mbox{\footnotesize fiber: 
}}{\gcd(n(k),n(t))\neq1}} \;\;
\sum_{\stackrel{\scriptstyle 0\leq l<N,}{l\equiv L\mbox{ \footnotesize  mod } n(t)}}
\vartheta^{l+kt} (E_k), \nonumber
\end{eqnarray}
where for $n(t)=2$ we set $E_t^{(1)}:=E_t$, and in the sum over
$k\in$ fiber each fixed point on the torus is counted separately.
As noted in Sect.~\ref{tori}, $\gamma_{MS}(T_0^{(\prime)})$ is a lattice
automorphism of order $4$, in fact a hyperk\a hler rotation. For our
extension (\ref{generalms}) of $\gamma_{MS}(T_0^{(\prime)})$ to the 
Kummer type lattice $\Pi_N$ generalizing $\Pi$ we find
\begin{eqnarray}\label{square}
&&\gamma_{MS}(X_0^{(\prime)})\circ \gamma_{MS}(X_0^{(\prime)})_{\mid \Pi_N}
= - \vartheta^{-1}\iota,\\
&&\quad\quad\quad\mbox{where for }
t\in I, l\in\{1,\dots,n(t)-1\}:\quad
\iota:\; E_t^{(l)}\longmapsto E_{-t}^{(l)}.\nonumber 
\end{eqnarray}
Hence $\gamma_{MS}(X_0^{(\prime)})$ has order $4,12,8,12$ for 
$N=2,3,4,6$, respectively.

For all fibers discussed above, the geometric action is analogous to
that on $I_0^\ast$: Up to   signs and possible corrections in 
$\pi_\ast H^{even}(T_0^{(\prime)},\Z)^{\Z_N}$ and
$\widehat\Pi_N$, the generalization of $\widehat\Pi$,   
$\gamma_{MS}(X_0^{(\prime)})$ exchanges the center node of 
the extended Dynkin diagram that describes the fiber with each component
of its long legs.

Of course, 
$\gamma_{MS}(X_0^{(\prime)})_{\mid\Pi_N}$
is not uniquely determined by the two conditions
$\gamma_{MS}(X_0^{(\prime)})$ $\in Aut(L(X_0^{(\prime)}))$ and
$\gamma_{MS}(X_0^{(\prime)})_{\mid \Pi_N^\perp} = \gamma_{MS}(T_0^{(\prime)})$.
In fact, assume that $\gamma_{MS}(X^{(\prime)}_0)g$ is another
consistent extension of $\gamma_{MS}(T_0^{(\prime)})$. Then by definition
$g_{\mid \Pi_N^\perp}=\id$, and $g$ acts as lattice automorphism both on
$L(X_0^{(\prime)})$ and $\Pi_N$. It is clear that $g$ can 
incorporate arbitrary shifts
of the fixed points in direction of the base in our fibration
$p:X\longrightarrow\P^1$. In case $N=2$ this corresponds to the freedom of
choice of $t_0\in I$ in (\ref{k3ms}). 
One checks that among the lattice automorphisms that permute fixed points,
$g$ must respect our fibration and therefore 
can only incorporate shifts on $I$ or 
the map $\widetilde\iota:t\mapsto-t, t\in I$. 
The latter is nontrivial only in the 
$\Z_3$ case $I\cong\F_3^2$, where it corresponds to the standard
$\Z_2$ action on the underlying torus, i.e.\ the algebraic automorphism 
that is modded out from $\widetilde{T_0^\prime/\Z_3}$ to obtain
$\widetilde{T_0^\prime/\Z_6}$. Having said this and
using the form of 
(\ref{generalms}), in the following we may
assume that $g$ acts as lattice automorphism 
on each $\E_t=\span_\Z\{ E_t^{(l)}, l\in\{1,\dots,n(t)-1\}\}$ separately.
Since $g$ is orthogonal, it must permute the roots in $\E_t$.  
Recall from \cite[Th. 3.3]{we00} that 
\begin{eqnarray}\label{upsilon0}
\widehat \upsilon^0 = \inv{\sqrt N}\upsilon - 
\inv{N} B_N + \widehat\upsilon,\\
\mbox{ where }  B_N\in\Pi_N: &&\forall\,t\in I,l\in\{1,\dots,n(t)-1\}:\;
\left\langle B_N, E_t^{(l)}\right\rangle = 1 . \nonumber
\end{eqnarray}
Then $g(\widehat E_t^{(l)})\in L(X_0^{(\prime)})$ with (\ref{ehat}) 
together with the fact that $g$ preserves scalar products implies
$g_{\mid \E_t} = \vartheta^{k_t}$ for some $k_t\in \{1,\dots,n(t)\}$.
Moreover, $g$ must obey
$$
g(\widehat\upsilon^0)-\widehat\upsilon^0=\inv{N} (1-g) B_N\in \Pi_N,
$$
which restricts the possible combinations of $k_t$ that define $g$. In fact,
with the results of \cite{we00} about the form of $\Pi_N$,
all in all we find that $\gamma_{MS}(X_0)$ is given by (\ref{generalms})
up to the above mentioned permutations in $I$ and the action of some
\begin{equation}\label{restrict}
g^b\in Aut(\Pi_N):\; g^b_{\mid\E_t}= \vartheta^{k_t}, \quad
g^b\left( \inv{N}  B_N\right) = \inv{N} B_N+b, \quad
b\in  \Pi_N\big/\bigoplus_{t\in I}\E_t.
\end{equation}
For $N\in\{2,3\}$ these degrees of freedom are parametrized by $Aff(I,\F_N)$,
since we have a natural identification
$$
g^b\quad\stackrel{1:1}{\longleftrightarrow}\quad k^b\in Aff(I,\F_N):
\quad g^b_{\mid\E_t}= \vartheta^{\,k^b(t)}.
$$
We will give an interpretation of this result at the end of Sect.~\ref{mckay}.
Note that by the above for any $a\in\Z$,
$\gamma_{MS}(X_0^{(\prime)})\vartheta^a=\vartheta^a\gamma_{MS}(X_0^{(\prime)})$ are
extensions of $\gamma_{MS}(T_0^{(\prime)})$ to $L(X_0^{(\prime)})$ as well.
Hence (\ref{square}) shows that
we can define mirror maps of order $4$ on the $\Z_3$ and $\Z_6$
orbifold limits of $K3$.

We can also easily check the action of $\gamma_{MS}(X_0^{(\prime)})$ on 
the monodromy
around the singular fibers. Again from \cite[Th.9.1]{ko63b} we  read
off the monodromy matrices and -apart from $I_0^\ast$-
find invariance under (\ref{mimo}) for $III^\ast$
type fibers only. It follows that the geometric interpretation as
$\Z_4$ orbifold limit obtained
from a toroidal theory on a rectangular torus with vanishing B-field is mapped
to another
such geometric interpretation under mirror symmetry, as expected. For the
$\Z_3$ and $\Z_6$ cases, on the other hand, the analogous statement is not
true. Though the conjugacy class of the monodromy remains unchanged under 
$\gamma_{MS}(X_0^{\prime})$ by definition, its representative changes under 
(\ref{mimo}). 
This might not be surprising because of the modification (\ref{tormsz3}) 
by a classical symmetry of order $4$ that we
had to perform on the hyperk\a hler rotation $\gamma_{MS}(T_0)$ to gain
our lattice automorphism $\gamma_{MS}(T_0^{\prime})$. It also
seems to be due to differences in the complex structure. Namely,
for rectangular tori one has a purely imaginary complex structure on the fiber,
but for $\Z_3$ or $\Z_6$ orbifolds this is impossible. The images under
mirror symmetry   will not have purely imaginary K\a hler parameters, in other
words will have a non-vanishing B-field. It might be interesting to explore
this point, which suggests a link between monodromy and complexified
K\a hler structure within classical geometry. In particular, it suggests
that there are interesting subtleties resulting from a transition from 
the universal covering space $\widetilde{\mathcal M}$ of our moduli space
to $\mathcal M$.
\section{The Mirror Map for $\Z_N$ Orbifold Conformal Field 
Theories}\label{twist}
By our discussion in the Introduction, $\gamma_{MS}(X_0^{(\prime)})$ 
is to be interpreted
as automorphism on the smooth universal cover $\widetilde{\mathcal M}$ 
of the moduli space
of $N=(4,4)$ superconformal field theories on $K3$ that identifies
equivalent theories. 
Let $m$ be a point in $\widetilde{\mathcal M}$ and $m^\prime$ its
image under this automorphism. Given a path between
$m$ and $m^\prime$ one can deform the CFT along this path. Since
$\widetilde{\mathcal M}$ is smooth and simply connected, 
the deformation (see e.g.\ \cite{ku89})
is defined up to an element of the holonomy group of the moduli space.
The quantum numbers of the fields, in particular the conformal dimensions
change continuously under deformations, but in general the deformation is
defined up to linear transformations among fields of identical quantum
numbers. When there are no degeneracies, this just amounts to an arbitrary
wave function renormalisation. When the conformal dimension of a field
changes along a path, the perturbation integrals for the change of its
$n$-point functions are logarithmically divergent and need a
regularisation, which introduces the arbitrariness just mentioned.
For chiral fields or their spectral flow partners, like the twist fields,
such a logarithmic divergence does not occur.
Their holonomy is trivial, too, as long as we deform within an
orbifold component of $\widetilde{\mathcal M}$. 
Since our version of mirror symmetry is induced by
T-duality on a $c=3$ toroidal subtheory of the underlying $c=6$ toroidal
superconformal field theory, we can determine the action of 
$\gamma_{MS}(X_0^{(\prime)})$
independently from the geometric results of Sects.~\ref{kummer} and 
\ref{kummertype}. This is the aim of the present section, where
it suffices to restrict to the bosonic subsector of each of our theories. 
We remark that
our technique is similar to that used in \cite{ber99}.

Orbifold CFTs  have a generic W-algebra given by the invariant part of
the current generated algebra on the underlying torus.
The Hilbert space of such a theory decomposes into the $\Z_N$ invariant
W-algebra representations of the underlying toroidal theory and the 
so--called twisted sector. The latter consists of
W-algebra representations with infinite
quantum dimensions, the ground states of which are given by the twist fields. 
In the following, these twist fields 
are denoted $T_t^l$, $l\in\{1,\dots,n(t)-1\}$,
where  we have chosen  $T_t$ of order $n(t)$ for each $t\in I$.  
We   normalize them such that
\begin{equation}\label{zamolodchikov}
\left\langle T_t^l, T_{t^\prime}^{l^\prime}\right\rangle
= \lim_{x\rightarrow0} |x|  
\left\langle T_t^l(x) T_{t^\prime}^{-l^\prime}(0)\right\rangle
= \delta_{t,t^\prime}\delta_{l,l^\prime} 
\left(2-\zeta_{n(t)}^l-\zeta_{n(t)}^{-l}\right).
\end{equation}
Here $\langle\cdot,\cdot\rangle$ denotes the standard scalar product on the 
Hilbert space which on $(1,1)$ fields induces the
Zamolodchikov metric.
In Sect.~\ref{mckay}, we will see that the normalization (\ref{zamolodchikov})
is the natural one.

Twist fields and all   the fields of the torus theory can
be included together in $n$-point functions, if one admits
ramifications of the world sheet at the twist field positions 
\cite{hava87,dfms87}. Denote by $V(p,z)$ the vertex operator 
of momentum $p=(p_l,p_r)$
and conformal dimensions $(h,\bar h)=(p_l^2/2,p_r^2/2)$,
where in the notations of Sect.~\ref{tori} we have 
$p=(p_l,p_r)={1\over\sqrt2}(Q_l,Q_r)$, i.e.\
${1\over\sqrt2}(Q_l+Q_r,Q_l-Q_r)\in v(\Lambda,B)\Z^{4,4}$.
The OPE of  $V(p,z)$
with the twist fields $T_t^l(x),\, t\in I$, has a contribution 
given by the twist fields themselves. We write it in the form
\begin{equation}\label{vaction}
V(p,z)T_t^l(x)= (z-x)^{-h} (\bar z- \bar x)^{-\bar h}
\!\!\!\!\!\!
\sum_{\stackrel{\scriptstyle t^\prime\in I,}{n(t)l^\prime=ln(t^\prime)}}
\!\!\!\!\!\!
W_{t t^\prime}(p)T_{t^\prime}^{l^\prime}(x) + \mbox{ other terms }.
\end{equation}
We will read off
the commutation relations of the matrices $W(p)$  
from the four-point function
$$
\begin{array}{rcl}
\langle V(p,z)V(p^\prime,z^\prime)T_t^l(x)T_{t^\prime}^{l^\prime}(y)
\rangle&=& 
e(z,z^\prime,x,y)O(z,z^\prime,x,y)\langle 
T_{t^\prime}^{-l^\prime}(x)T_{t^\prime}^{l^\prime}(y)\rangle,\\[4pt]
e(z,z^\prime,x,y)&:=&(z-x)^{-h} (\bar z- \bar x)^{-\bar h} 
(z^\prime-x)^{-h^\prime} (\bar z^\prime- \bar x)^{-\bar h^\prime} \\[4pt]
&&\;(z-y)^{-h} (\bar z- \bar y)^{-\bar h} 
(z^\prime-y)^{-h^\prime} (\bar z^\prime- \bar y)^{-\bar h^\prime} 
. 
\end{array}
$$
The function $O$ can be written as the product of two functions which
are analytic resp. anti-analytic in their corresponding domains.

Let us assume that $T_t^l,\,T_{t^\prime}^{l^\prime}$ 
both correspond to $\Z_n$ twists
$\theta$, in particular
$$
n=n(t)/\gcd(l,n(t))=n(t^\prime)/\gcd(l^\prime, n(t^\prime)),
$$
such that  by moving $z$ in a loop 
around $x$ in $T_t^l(x)$ or around $y$ in $T_{t^\prime}^{l^\prime}(y)$ 
in the opposite sense,
$V(p,z)$ becomes $V(\theta p,z)$. 
Hence the $V(p,z)$ are well defined on
the $n$-fold cover of the Riemann sphere, which is another Riemann sphere
with coordinate $\xi=((z-x)/(z-y))^{1/n}$, and
analogously $\eta=((z^\prime-x)/(z^\prime-y))^{1/n}$.

For fixed $x,y$ and the appropriate domains for $z,z^\prime$,
the function $O(z,z^\prime,x,y)$ is analytic on the cover, 
with
poles given by the known OPE of the vertex operators.
This yields
$$
O(z,z^\prime,x,y)=o(x-y)
\prod_{k=1}^n (\xi-\zeta_n^k\eta)^{p_l\theta^k p^\prime_l}
(\bar\xi-\zeta_n^{-k}\bar\eta)^{p_r\theta^k p^\prime_r} ,
$$
where $\zeta_n:=\exp(2\pi i/n)$ and $o(x-y)$ is easy to calculate
but irrelevant for our purpose. Taking the limits
$z\rightarrow x$ and $z^\prime\rightarrow x$ in different orders
one finds
\begin{equation}\label{weyl}
\begin{array}{rcl}
W(p^\prime)W(p)&=&\ds
\prod_{k=1}^n \zeta_n^{k p\theta^k p^\prime}W(p)W(p^\prime)
= \zeta_n^{\phi_n(p,p^\prime)} W(p)W(p^\prime) ,\\[-3pt]
\mbox{where }\hphantom{\phi(p,p^\prime)} \\[-3pt]
\phi_n(p,p^\prime)&:=&  \ds\sum_{k=1}^n k p\theta^k p^\prime
\;\mbox{ and }\; qq^\prime\stackrel{(\ref{bil})}{=}
q_lq^\prime_l-q_r q^\prime_r
= Q_lQ_r^\prime+Q_rQ_l^\prime 
\end{array}
\end{equation}
(see \cite{le85,nsv87}).
Hence the $W(p)$ form a Weyl algebra which is represented on the
vector space spanned by the twist fields. 

For a given
geometric interpretation, where we  assume 
$B=0$ on the underlying toroidal theory,
the momentum state vertex operators
are characterized by $p_l=p_r$. They therefore
form an Abelian subalgebra of the
Weyl algebra (\ref{weyl}). With respect to this subalgebra, the
representation decomposes into one-dimensional subrepresentations
with ground states $T_t^l$,
each of which corresponds to a $\Z_{n(t)/\gcd(l,n(t))}$ twist on a
$\Z_{n(t)}$ type fixed point. 

To see this explicitly   
note that for a geometric interpretation with $B=0$ we have
\begin{eqnarray*}
&&p=(p_l,p_r)=\inv{\sqrt2}(\mu+\lambda,\mu-\lambda)=:p(\mu,\lambda)\\
&&\quad\quad
\mbox{ with } \quad (\mu,\lambda)\in\span_\Z\{\mu_1,\dots,\mu_4\}\oplus
\span_\Z\{\lambda_1,\dots,\lambda_4\}.
\end{eqnarray*}
Now (\ref{weyl}) takes the form
\begin{equation}\label{commu}
\phi_n(p,p^\prime) 
\equiv  \sum_{k=1}^n k
\left(\mu\theta^k\lambda^\prime-\mu^\prime\theta^k\lambda\right)
\mod n\Z.
\end{equation}
In fact, the fixed point set
$I$ can be interpreted as part of the  subgroup
of elements of order $n$ with $n\!\mid\!N$ in the Jacobian torus
$H_1(T_0,\R)/H_1(T_0,\Z)$, where
$H_1(T_0,\Z)\cong\span_\Z\{\lambda_1,\dots,\lambda_4\}$, 
and we can use
$$
I \hookrightarrow H_1(T_0,\Q)/H_1(T_0,\Z), \quad
\mbox{ where } n\lambda\in \span_\Z\{\lambda_1,\dots,\lambda_4\}
\mbox{ for } [\lambda]\in I.
$$
Then (\ref{vaction}) reads, after suitable normalizations of the vertex
operators,
\begin{eqnarray}\label{rep}
V(p(\mu,\lambda),z)T_t^l(x)
= (z-x)^{-h} (\bar z- \bar x)^{-\bar h}
&&\!\!\!\!\!\!
\zeta_{n(t)}^{l\mu(nt)}T_{t^\prime}^{ln(t^\prime)/n(t)}(x) + \mbox{ other terms },
\nonumber\\
t^\prime:=t+\left[(1-\theta)^{-1}\lambda\right]
&=& t-\left[\inv{n(t)}\smash{\sum_{k=1}^{n(t)-1}}k\theta^k\lambda\right],
\vphantom{\sum_{k=1}^{n(t)}}\\
\theta^n&=&\id,\quad\quad 
n=\gcd(l,n(t)),\nonumber\\[3pt]
\mbox{ which indeed yields }\quad\quad\quad
W(p^\prime)W(p)T_t^l
&=& \zeta_{n(t)}^{\;l\phi_{n}(p,p^\prime)} W(p)W(p^\prime)T_t^l .\nonumber
\end{eqnarray}
Now recall that $\gamma_{MS}(T_0^{(\prime)})$ as 
discussed in Sect.~\ref{tori} is given by
T-duality on the $x_1,x_2$ directions of the torus $T$, which agrees with
the Fourier-Mukai transform on the corresponding two-torus 
\cite{na82,na84,sc88,brvb89}, see also \cite{th01}. 
The standard Fourier transform on $\R^d$
extends to all measurable Abelian groups. On
$\Z_N$ it takes the form 
$$
\widetilde F^{(N)}:\C^N\longrightarrow\C^N, \quad
\widetilde F^{(N)}\left(\vphantom{\sum}(f_k)_{k=1}^N\right) = 
\left(\inv{\sqrt N}\smash{\sum_{j=1}^{N}}
\vphantom{\sum} f_j \;\zeta_N^{jk}\right)_{k=1}^N.
\vphantom{\sum_{j=1}^{N}}
$$
In the present context, however, it is natural to restrict $\widetilde F^{(N)}$
to $\Z_N\subset U(1)\subset\C^\ast$ on each component. Let $t_k\in\Z_N$ denote
a generator in the $k^{th}$ component, then one has
$$
 F^{(N)}(t_k^l)  = \inv{\sqrt N}\sum_{j=1}^{N} t_j^{l} \;\zeta_N^{jkl},
$$
which for simplicity we keep on calling discrete Fourier transform
in the following.

Under T-duality, momentum states and winding states are
interchanged, and the latter are characterized by $p_l=-p_r$.
The ground states of the one-dimensional subrepresentations 
of the subalgebra of winding
states are therefore obtained from the generators $\{T_t^l,t\in I,
l\in\{1,\dots,n(t)-1\}\}$ of the twisted sector
by performing a $\Z_N$ type discrete Fourier  transform. By the above,
the resulting map on twist fields has the general form
\begin{eqnarray}\label{ftwist}
l\in\{1,\dots,n(t)-1\}:\nonumber\\
F(T_t^l) 
&=& \sum_{\stackrel{\scriptstyle j\in\mbox{ \footnotesize fiber: 
}}{n(t)\mid ln(j)}}
  a^l_{n(j),n(t)} T_j^{ln(j)/n(t)}\zeta_{n(t)}^{ltj},  \\
\quad\quad\mbox{ with } \forall\;\lambda\in\C:\quad
F(\lambda T_t^l) &=& \lambda F(  T_t^l)
\nonumber
\end{eqnarray}
and the coefficients $a^l_{n(j),n(t)}$ to be determined.
Here, we use the labeling of fixed points that was introduced in 
Sect.~\ref{kummertype} and which is restricted to the
fiber coordinates $x_1,x_2$. As in (\ref{generalms}), each fixed point
on the torus contributes separately to the sum $j\in$ fiber.

When $N$ factorizes, the orbifolding can be done stepwise.
Therefore,
$$ 
a^{kl}_{n,kn^\prime} = k^{-1/2} a^l_{n,n^\prime},
\quad\mbox{ and }\quad
a^l_{kn,n^\prime} = k^{1/2} a^l_{n,n^\prime}.
$$
This leaves only $a^1_{n,n}$ to be determined. Since $F$
must conserve the normalization, counting fixed points gives
$$
|a^1_{n,n}|^{-2} = 2-\zeta_n-\zeta_n^{-1}.
$$
Up to a finite ambiguity, the phase can be determined from the
order of $F$. Instead of a cumbersome direct determination, we
fix it by consistency requirements with the
results of Sects.~\ref{kummer},\ref{kummertype} (see
Sect.~\ref{mckay}). This yields:
$$
a^l_{n,n^\prime}
=- \inv{\sqrt{n n^\prime}}\sum_{m=1}^{n-1} m \zeta_{n^\prime}^{lm}
= \sqrt{\inv[n\,]{n^\prime}} (1-\zeta_{n^\prime}^{l})^{-1}.
$$
Note in particular that
$$
F\left(\overline{T_t^l}\right)=F\left(T_t^{n(t)-l}\right)=\overline{F(T_t^l)}.
$$
Explicitly we have in the $\Z_2$ case:
$$
t\in\F_2^2:\quad F(T_t) = \inv{2} \sum_{j\in\F_2^2} (-1)^{tj}T_j,
$$
in the $\Z_3$ case:
$$
t\in\F_3 :\quad 
F(T_t) =  -\inv[i\omega]{\sqrt3} \sum_{j\in\F_3} \omega^{tj}T_j,\;\;
\omega=-\inv{2}+i\inv[\sqrt3]{2},
$$
in the $\Z_4$ case:
\begin{eqnarray*}
t\in\F_2   :\quad 
F(T_{(t,t)}) &=& \inv[1+i]{2} 
\left(T_{(0,0)} + (-1)^tT_{(1,1)}\right),\\
F(T_{(t,t)}^2) &=& \inv{2} \left( \sqrt2 (-1)^t T_{(0,1)}+ 
T_{(0,0)}^2 + T_{(1,1)}^2\right),\\
F(T_{(0,1)}) &=& \inv{\sqrt2} \left(T_{(0,0)}^2 -T_{(1,1)}^2\right),
\end{eqnarray*}
in the $\Z_6$ case:
$$
\begin{array}{rclrcl}
F(T_0) &=&  \widetilde\omega\; T_0,\;\;
&\widetilde\omega&=&\inv{2}+i\inv[\sqrt3]{2},\\[6pt]
F(T_0^2)&=& -\inv[i\omega]{\sqrt3} 
\left(T_0^2 +  \sqrt2 T_1\right),\;\;
&\omega&=&-\inv{2}+i\inv[\sqrt3]{2},\\[6pt]
F(T_0^3) &=& \inv{2} 
\left(T_0^3 +  \sqrt3 T_{(1,0)}\right),\\[6pt]
F(T_1) &=&  \inv[i\omega]{\sqrt3} 
\left(T_1 - \sqrt2 T_0^2  \right), \\[6pt]
F(T_{(1,0)})&=& -\inv{2} \left(T_{(1,0)}-\sqrt3 T_0^3   \right).
\end{array}
$$
\section{Mirror Symmetry,   Discrete Fourier  Transform, and
the McKay Correspondence}\label{mckay}
In Sects.~\ref{kummertype} and \ref{twist} we have independently
derived the action of mirror symmetry $\gamma_{MS}(X_0^{(\prime)})$ 
both in terms of
geometric data (\ref{generalms}) and CFT data (\ref{ftwist}).
Now we shall relate the two approaches and give an interpretation 
in terms of the classical McKay correspondence.

In general, we assume that the data of the
$\Z_N$ orbifold CFT as discussed in 
Sect.~\ref{twist} possess a geometric interpretation on our 
$\Z_N$ orbifold limit of $K3$ as discussed in Sect.~\ref{kummertype}.
In view of our results on mirror symmetry, we may expect a linear map
that intertwines between the pictures. More precisely, on the CFT side
the twist fields $T_t^l$
generate deformations of the theory, which
act linearly on the   cycles Poincar\'e dual to the
$E_t^{(l)}$. Thus they should 
naturally correspond to linear combinations
of these cocycles. We denote the induced linear map from
cocycles to twist fields by $C$. This map should  act isometrically
with respect to the standard metrics (\ref{zamolodchikov}) and
(\ref{intersection})  
up to a global sign, and it should obey
\begin{equation}\label{intertwine}
\forall\; t\in I, l\in\{1,\dots,n(t)-1\}:\quad
F\;C( E_t^{(l)} )\;\;=\;\;C\;\gamma_{MS}(X_0^{(\prime)})( E_t^{(l)} ).
\end{equation}
On the CFT level, the sectors of a $\Z_N$ orbifold theory
carry distinct representations of the dual of $\Z_N$, with the
untwisted sector corresponding to the trivial representation.
In particular, one obtains nontrivial representations on all
twist fields. More precisely, we have 
an action of the dual of $\Z_{n(t)}$ on the twist fields
$T_t^l$ at a given fixed point $t\in I$. In Sect.~\ref{twist} 
we have already labeled the twist fields such that the action is 
generated by multiplication with $\zeta_{n(t)}^l$ on $T_t^l$.
By abuse of notation we denote the generator of this $\Z_{n(t)}$ 
action by $\vartheta$. It gives the ``quantum symmetry'' of the 
orbifold conformal field theory that replaces the ``geometric
symmetry'' on the underlying toroidal theory. 
It is well known that by modding out   the orbifold
CFT by the total ``quantum'' $\Z_N$ symmetry we can
reproduce the original
toroidal theory.

Geometrically, each fixed point $t\in I$ is left invariant by a
subgroup $\Z_{n(t)}$ of $\Z_N$. This
induces an action of the dual of $\Z_{n(t)}$ on 
$\E_t=\span_\Z\{E_t^{(l)}, l\in\{0,\dots,n(t)-1\}\}$. Indeed,
its generator $\vartheta$ has been introduced in (\ref{thetadef}) above.
The map $C$ can therefore be obtained
by decomposing the integral $\Z_{n(t)}$ action on  $\E_t$ 
into one-dimensional representations.
We claim that this yields
\begin{equation}\label{trafo}
t\in I, l\in\{1,\dots,n(t)-1\}:\quad
C( E_t^{(l)} ):=
\inv[1]{\sqrt{n(t)}}
\sum_{k=1}^{n(t)-1}  \zeta_{n(t)}^{lk} T_t^k.
\end{equation}
Together with (\ref{generalms}), (\ref{ftwist}) one 
first checks (\ref{intertwine}).
Moreover, the reality condition
$$
\forall\; t\in I, l\in\{1,\dots,n(t)-1\}:\quad
\overline{C ( E_t^{(l)} )}\;\;=\;\;C( E_t^{(l)} ),
$$
translates into $\overline{T_t^l}\;\;=\;\;T_t^{n(t)-l}$, as
expected.

Finally, up
to a prefactor, the Hermitean form on $\E_t$ is given
by the intersection product (\ref{intersection})
on the cocycles, thus by the Cartan matrix of the extended Dynkin diagram. 
Since the latter can be written as $2-\vartheta-\vartheta^{-1}$, 
it commutes with $\vartheta$ and becomes diagonal in the basis $T_t^l$ 
of eigenvectors of $\vartheta$. 
With $C$   a unitary   matrix  with respect to this basis,
the squared norms of the $T_t^l$ yield the corresponding eigenvalues
$2-\zeta_{n(t)}^l-\zeta_{n(t)}^{-l}$ of the Cartan matrix, in agreement
with (\ref{zamolodchikov}). All in all, (\ref{trafo}) is checked to give
an isometry, up to a global sign, on the entire twisted sector.

Now (\ref{trafo}) shows that (\ref{rep}) in fact induces an action of
$\span_\Z\{\mu_1,\dots,\mu_4\} \oplus \span_\Z\{\lambda_1,\dots,\lambda_4\}$
on $\Pi_N$. The action of $\span_\Z\{\mu_1,\dots,\mu_4\}$ is given by
\begin{equation}\label{nextrep}
\begin{array}{rcl}
\forall\;\mu\in\span_\Z\{\mu_1,\dots,\,\mu_4\} 
,\;
t\in I,&&  l\in\{1,\dots,n(t)\!-\!1\}:\\[5pt]
W(\mu,0)E_t^{(l)} &:=& E_{t}^{(\mu(nt)+l)}
\stackrel{(\ref{thetadef})}{=} \vartheta^{\mu(nt)}E_{t}^{(l)}.
\end{array}
\end{equation}
The action of $\span_\Z\{\lambda_1,\dots,\lambda_4\}$ is more complicated,
at least in this basis. It translates directly into a natural action
on the basis
$$
t\in I, l\in\{1,\dots,n(t)-1\}:\quad
C^{-1}(T_t^k) 
\stackrel{(\ref{trafo})}{=}
{1\over\sqrt{n(t)}}
\sum_{l=0}^{n(t)-1} \zeta_{n(t)}^{lk} E_t^{(l)},
$$
along with the corresponding Weyl algebra representation given
in Sect.~\ref{twist}. In order to work with geometric objects, recall
from our discussion in Sect.~\ref{kummertype} that rather than 
the lattice $\Pi_N$
we have to consider its projection $\widehat\Pi_N$ onto
$H^2(X_0^{(\prime)},\Z)$. By $\widehat\upsilon^0,\widehat\upsilon$ we denote the
generators of $H^0(X_0^{(\prime)},\Z),H^4(X_0^{(\prime)},\Z)$, respectively, 
and recall from
(\ref{ehat}) that  
$$
\forall\,
t\in I, l\in\{1,\dots,n(t)-1\}:\quad
E_t^{(l)}=\widehat{E}_t^{(l)}-\inv{n(t)}\widehat\upsilon.
$$ 
Hence the $\Z_{n(t)}$ symmetry (\ref{thetadef})   on
the irreducible components of the
exceptional divisor over a given fixed point $t\in I$ actually is    
$$
\vartheta\left(\widehat{E}_t^{(l)}\right) = \left\{ 
\begin{array}{ll}
\widehat{E}_t^{(l+1)} & \mbox{ if } l<n-1,\\
\widehat\upsilon
-\sum\limits_{j=1}^{n-1}\widehat{E}_t^{(j)} & \mbox{ if } l=n-1,
\end{array}\right.
$$
i.e.\ $\vartheta(\widehat{\upsilon})=\widehat{\upsilon}$. 
By the above, the ``quantum'' symmetry $\vartheta$   has
a straightforward geometric meaning.

We also set
$\vartheta(\widehat{\upsilon}^0)=\widehat{\upsilon}^0$ and consider the induced
action on the basis dual to 
$\{\widehat{E}_t^{(l)}, l\in\{1,\dots,n(t)-1\},\widehat\upsilon
-\sum\limits_{j=1}^{n-1}\widehat{E}_t^{(j)}\}$:
With $\{ (\widehat{E}_t^{(l)})^\ast \} \subset 
\span_\Q\{ \widehat{E}_t^{(l)} \}$ the dual basis with respect to 
the fundamental system $\{ \widehat{E}_t^{(l)} \}$ of
$\widehat{\E}_t=\span_\Z\{ \widehat{E}_t^{(l)} \}$ let
\begin{equation}\label{mukai}
\varepsilon_t^{(l)}
:= 
\left\{
\begin{array}{lcl}
\widehat{\upsilon}^0 &\mbox{ if }& l=0,\\
\widehat{\upsilon}^0 + (\widehat{E}_t^{(l)})^\ast
&\mbox{ if }& 0<l<n(t)
\end{array}
\right\},   
\quad\quad
\forall\; l\in\Z:\; \varepsilon_t^{(l+n(t))} = \varepsilon_t^{(l)}.
\end{equation}
By (\ref{nextrep}),
$\{\varepsilon^{(0)}, \dots, \varepsilon^{(n(t)-1)}\}$
carries the representation analogous to (\ref{nextrep}), which obeys
\begin{eqnarray}\label{finalrep}
\mu\in
\span_\Z\{\mu_1,\dots,\mu_4\} ,&&
t\in I,\; l\in\{1,\dots,n(t)-1\}:\nonumber\\[5pt]
W(\mu,0)\varepsilon_t^{(l)} 
&=& \varepsilon_{t}^{(\mu(nt)+l)},\quad n=n(t), \\[5pt]
W(\mu,\lambda)W(\mu^\prime,\lambda^\prime)
&=& \vartheta^{\phi_{n}(p,p^\prime)}
W(\mu^\prime,\lambda^\prime)W(\mu,\lambda),\;
p=p(\mu,\lambda),\;p^\prime=p(\mu^\prime,\lambda^\prime)\nonumber
\end{eqnarray}
with $\phi_{n}(p,p^\prime)$ as in (\ref{commu}).

This is readily interpreted in terms of $\Z_N$ equivariant topological
K-theory on the underlying torus $T$. Namely, consider a $\Z_N$ equivariant
flat line bundle $\mathcal L$ on $T$, then by \cite{bkr99} there is a 
corresponding line bundle $\pi_\ast\mathcal L$ on our $K3$ surface
$X=\widetilde{T/\Z_N}$. More precisely, $\mathcal L$ is determined
by the representation $\chi_t$ of $\Z_N$ in each fiber over a fixed
point $t\in I$. Then 
$$
c_1(\pi_\ast {\mathcal L}) = \sum_{t\in I} c_1(\chi_t),
$$
where $c_1(\chi_t)$ is given by the classical McKay correspondence
\cite{mk80,mk81,gsve83,kn85}:
$$
c_1(\chi_t^{(l)})=(\widehat{E}_t^{(l)})^\ast, \quad
\mbox{ if } \chi_t^{(l)}:\C\longrightarrow\C,\quad
\chi_t^{(l)}(z)=\zeta_{n(t)}^l \; z.
$$
Hence $\varepsilon_t^{(l)}, t\in I, l\in\{0,\dots,n(t)-1\}$ 
as in (\ref{mukai}) are the $H^0(X,\Z)\oplus H^2(X,\Q)$ components
of the Mukai vectors
$$
ch({\mathcal E}) \sqrt{\widehat A(X)}
= [\mbox{rk}\mathcal E]\, \widehat\upsilon^0 +c_1(\mathcal E) 
+ \left[ (c_2-\inv{2}c_1^2)(\mathcal E)[X] + 
\inv[\mbox{rk}\mathcal E]{48} p_1(X) \right]
\widehat\upsilon
$$
for all possible   contributions $\mathcal E$ to $\pi_\ast\mathcal L$
over $t\in I$. The coefficient of $\widehat\upsilon $ does not enter into
the relation between line bundles and representations.

We therefore find that the action (\ref{finalrep}) of 
$\span_\Z\{\lambda_1,\dots,\lambda_4\}\cong H_1(T_0^{(\prime)},\Z)$ agrees with
a non-standard action  of the subgroup 
$I\subset H_1(T_0^{(\prime)},\R)/H_1(T_0^{(\prime)},\Z)$ of the Jacobian
torus on $\Z_N$ equivariant flat line bundles on $T$. It looks natural
in the dual of the basis $C^{-1}(T_t^k)$.
Moreover, $\mu\in\span_\Z\{\mu_1,\dots,\mu_4\}$
acts by tensoring a line bundle on $T$  with the
bundle $\mathcal L_\mu$ associated to $\chi$, where
$$
z\in (\mathcal L_\mu)_t, t\in I:\quad \chi(z)=\zeta_N^{\mu(n(t)t)}z.
$$
Summarizing, (\ref{trafo}) gives an isometry between 
the $\C$-vector space spanned by the twist fields of 
an orbifold CFT and a subspace of $H^\ast(X_0^{(\prime)},\C)$, 
which allows us to 
identify the natural action of the Weyl algebra (\ref{weyl}) on 
twist fields with an action of the same Weyl algebra on the
$\Z_N$ equivariant flat line bundles on $T$. It is interesting
to note that the action of the Jacobian torus is rather non-trivial.
It has been known before, of course,  that such identifications
should be possible \cite{wi98,gc99,digo00,bgk01,bdhkmms01}.

Recall from (\ref{restrict}) that our formula (\ref{generalms}) was
only determined up to certain permutations of the fixed points and
an action of 
$g^b$ that is characterized by inducing an integral B--field shift by some
$b\in \Pi_N/\bigoplus\limits_{t\in I}\E_t$. Since 
$\Pi_N/\bigoplus\limits_{t\in I}\E_t$, and for $N\in\{2,3\}$ equivalently
$Aff(I,\F_N)$
parametrizes $\Z_N$ equivariant flat
line bundles on $T$, this freedom of choice translates nicely into
possible  choices of the origin in that parameter space.
We have seen, though, that the choice of $g^b$   
may influence the order of the resulting  
$\gamma_{MS}(X_0^\prime)$. We find that it then corresponds to a twisting
of our mirror map, given by some equivariant flat
line bundle on $T$.

Finally, let us briefly explain the connection of our results to 
Chen and Ruan's orbifold cohomology \cite{chru00} on $[T/G]$ as 
discussed in \cite{ru00,fago01}. In the present case of cyclic
$G=\Z_N$ it is isomorphic to 
$$
H^\ast_{CR}([T/\Z_N],\C)
\cong H^\ast(T,\C)^{\Z_N}\oplus
\bigoplus_{t\in I} \bigoplus_{l=1}^{n(t)-1}
\C\mathcal T(t)_{\zeta_{n(t)}^l},
$$
where the generators $\mathcal T(t)_{\zeta_{n(t)}^l}$ of the
twisted sector corresponding to the fixed point $t\in I$ have 
scalar product
\begin{equation}\label{chenruan}
\left( \mathcal T(t)_{\zeta_{n(t)}^l}, 
\mathcal T(t^\prime)_{\zeta_{n(t^\prime)}^{l^\prime}}\right)
:= \delta_{t,t^\prime}\delta_{l,-l^\prime}.
\end{equation}
In fact, if instead of (\ref{chenruan})
one uses the associated sesquilinear form\footnote{\label{ruan2}As
remarked in footnote \ref{ruan}, this was explained to us by Yongbin Ruan
\cite{ru01},
and goes back to earlier observations by Edward Witten.} and slightly
different normalizations to identify $\mathcal T(t)_{\zeta_{n(t)}^l}$
with $T_t^l$, it agrees with our metric in 
the twisted sector of the orbifold CFT. More importantly, this means
that (\ref{trafo}) can be used to prove that for all orbifold limits 
discussed in 
this note
$$
H^\ast_{CR}([T_0^{(\prime)}/\Z_N],\C)\cong H^\ast(X_0^{(\prime)},\C)
$$
as metric spaces, confirming part of Ruan's conjecture 
\cite[Conj.6.3]{ru00} for these cases.
\section{Conclusions}\label{conc}
In this note, we have analyzed a version of mirror symmetry on elliptically
fibered $K3$ surfaces that is induced by fiberwise T-duality on nonsingular
fibers. It is straightforward to determine this map for four-tori, which
enables us to give the explicit action on $\Z_N$ orbifold limits of $K3$,
$N\in\{2,3,4,6\}$, by the techniques of \cite{nawe00,we00}. 
In the present case of $N=(4,4)$ superconformal field theories with central 
charge $c=6$ the mirror map can be realized as automorphism on the
lattice of integral cohomology on the underlying complex surface 
(torus or $K3$). While the order of mirror symmetry on the torus is $4$,
it can take values $4,8,$ or $12$ on our orbifold limits of $K3$.
On the CFT side, the mirror map
is induced by a $\Z_N$ type discrete Fourier  transform in the fiber
acting on the twist fields of the orbifold conformal field theory. Since
we are able to derive the mirror map in both the geometric and 
conformal field theoretic description independently, we can deduce the
exact geometric counterparts of orbifold CFT twist fields. Moreover,
the correspondence between the twist fields and   $\Z_N$ equivariant
flat line bundles can be deduced directly. The natural ``quantum'' $\Z_N$
symmetry in the twisted sector of the orbifold CFT, which can be modded
out to retain the original toroidal theory, gains geometric meaning. 
In fact, by the classical McKay correspondence it can be traced back to
properties of singularities already investigated in \cite{mu61,hi63}.

Our version of mirror symmetry agrees with the one proven more
generally in \cite{vawi95}, which was generalized to the celebrated
Strominger/Yau/Zaslow conjecture  \cite{syz96}. We have   avoided M-theory 
language, though, and restricted considerations to
the underlying geometry. Together with the rich structure of
the $K3$ moduli space this enables us to carry out the construction away
from large volume or large complex structure limits in the moduli space
and even without touching the issue of its
compactification.  Note that all
our $T^2$ fibrations $p:X\longrightarrow\P^1$ have non-stable singular 
fibers. A large complex structure limit in the sense of \cite{grwi00} has
therefore not even been defined for those cases we are interested in.

Our results on the explicit prescription for the identification of
conformal field theoretic and geometric data might be of interest in 
their own right. We have pointed out how they resolve the objection in
\cite{fago01} to Ruan's conjecture 
\cite[Conj.6.3]{ru00}
(see footnotes \ref{ruan} and \ref{ruan2};
\cite{ru01}).
Note also that we are working within the full CFT, without having to
perform a topological twist or introduce boundary states to probe the
geometry of our orbifold limits of $K3$.
%
%
\newcommand{\etalchar}[1]{$^{#1}$}
\def\polhk#1{\setbox0=\hbox{#1}{\ooalign{\hidewidth
  \lower1.5ex\hbox{`}\hidewidth\crcr\unhbox0}}}
\providecommand{\bysame}{\leavevmode\hbox to3em{\hrulefill}\thinspace}

\end{document}